\documentclass[journal = ancac3,manuscript = article]{achemso} 
\usepackage{color}

\usepackage{chemformula} 
\usepackage[T1]{fontenc} 
\usepackage[version=3]{mhchem} 
\setkeys{acs}{maxauthors=20, etalmode=truncate}

\author{Xiao Xiong}
\email{xiong_xiao@ihpc.a-star.edu.sg}
\affiliation{
Institute of High Performance Computing, A*STAR (Agency for Science, Technology and Research), 1 Fusionopolis Way, \#16-16 Connexis, Singapore 138632, Singapore.
}
\altaffiliation{These authors contribute equally.}%

\author{Yiming Lai}
\affiliation{
School of Physics and CRANN Institute, Trinity College Dublin, Dublin 2, Ireland.
}
\alsoaffiliation{ 
The Blackett Laboratory, Imperial College London, Prince Consort Road, London SW7~2AZ, United Kingdom.
}
\altaffiliation{These authors contribute equally.}%

\author{Daniel Clarke}%
\affiliation{ 
School of Physics and CRANN Institute, Trinity College Dublin, Dublin 2, Ireland.
}
\altaffiliation{These authors contribute equally.}%

\author{Nuttawut Kongsuwan}
\affiliation{ 
The Blackett Laboratory, Imperial College London, Prince Consort Road, London SW7~2AZ, United Kingdom.
}
\alsoaffiliation{%
Quantum Technology Foundation (Thailand), 98 Soi Ari, Bangkok, 10110, Thailand.
}%
\alsoaffiliation{%
Thailand Center of Excellence in Physics, Ministry of Higher Education, Science, Research and Innovation, Bangkok 10400, Thailand.
}%

\author{Zhaogang Dong}
\affiliation{%
Institute of Materials Research and Engineering, A*STAR (Agency for Science, Technology and Research), 2 Fusionopolis Way, \#08-03 Innovis, 138634 Singapore.
}%

\author{Ping Bai}
\affiliation{
Institute of High Performance Computing, A*STAR (Agency for Science, Technology and Research), 1 Fusionopolis Way, \#16-16 Connexis, Singapore 138632, Singapore.
}

\author{Ching Eng Png}
\affiliation{
Institute of High Performance Computing, A*STAR (Agency for Science, Technology and Research), 1 Fusionopolis Way, \#16-16 Connexis, Singapore 138632, Singapore.
}%

\author{Ortwin Hess}%
\email{ortwin.hess@tcd.ie}
\affiliation{ 
School of Physics and CRANN Institute, Trinity College Dublin, Dublin 2, Ireland.
}%
\alsoaffiliation{ 
The Blackett Laboratory, Imperial College London, Prince Consort Road, London SW7~2AZ, United Kingdom.
}

\author{Lin Wu}%
\email{lin_wu@sutd.edu.sg}
\affiliation{
Science, Mathematics and Technology (SMT), Singapore University of Technology and Design (SUTD), 8 Somapah Road, Singapore 487372.
}
\alsoaffiliation{
Institute of High Performance Computing, A*STAR (Agency for Science, Technology and Research), 1 Fusionopolis Way, \#16-16 Connexis, Singapore 138632, Singapore.
}%

\title{Controlling plexcitonic strong coupling via multidimensional hotspot nanoengineering}

\begin{document}


\begin{abstract}

Plexcitonic strong coupling has ushered in an era of room-temperature quantum electrodynamics that is achievable at the nanoscale, with potential applications ranging from high-precision single-molecule spectroscopy to quantum technologies functional under ambient conditions. Realizing these applications on an industrial scale requires scalable and mass-producible plasmonic cavities that provide ease of access and control for quantum emitters.
Via a rational selection of substrates and the canonical gold bowtie nanoantenna, we propose a novel design strategy for multidimensional engineering of nanocavity antenna-mode hotspots, which facilitates their elevation to the top of the nanobowtie gap and provides a field enhancement of $\sim500$ fold (a 1.6-fold increase compared to a conventional nanobowtie-on-glass cavity at the bottom of the nanobowtie gap). We discuss the formation mechanism for such antenna modes using different material substrates from the perspective of charge carrier motion, and analyze their sensitivity to the geometrical parameters of the device. The advantages of these antenna modes, particularly in view of their dominantly in-plane polarized near-fields, are further elaborated in a spatiotemporal study of plexcitonic strong coupling involving single emitters and layered ensembles thereof, which reveals ultrafast quantum dynamics dependent on both the substrate and nanobowtie geometry, as well as the potential for applications related to 2D materials whose excitonic dipoles are typically oriented in-plane. 
The conceptual discovery of this substrate-enabled antenna-mode nanoengineering could readily be extended to tailor hotspots in other plasmonic platforms, and we anticipate that this work could inspire a wide range of novel research directions from photoluminescence spectroscopy and sensing to the design of quantum logic gates and systems for long-range energy transfer.

\end{abstract}


\section{I. Introduction}

Realizing strong light-matter interaction with a single quantum emitter has been an essential ingredient of many quantum technologies \cite{xu2018quantum} including optoelectronics \cite{schweicher2020molecular}, polariton chemistry \cite{ribeiro2018polariton}, and ultrafast spectroscopy \cite{silva2020polaritonic}. Although this phenomenon was previously exclusive to experiments at cryogenic temperatures, recent advances in plasmonics have brought light-matter interaction to the era of room-temperature single-emitter strong coupling \cite{chikkaraddy2016single,liu2017strong,Grossetal2018}. The intense electromagnetic field confinement provided by plasmonic nanocavities compensates for the severe Ohmic loss in metals and enables rapid energy exchange between plasmon polaritons and matter excitons, giving rise to hybrid plasmon-excitons or plexcitons \cite{hensen2018strong,xiong2020ultrastrong,xiong2021room}. Strongly-coupled plexcitonic systems allow for a wide range of applications including ultrafast single-photon emission \cite{bogdanov2020ultrafast}, single-qubit coherent control \cite{zhang2017sub}, charge transport \cite{orgiu2015conductivity}, long-range energy transfer \cite{bouchet2016long,aeschlimann2017cavity}, and universal quantum logic gates \cite{calafell2019quantum}, which provide fundamental building blocks for optical communication and quantum information processing. In the context of polariton chemistry, plexcitonic systems can efficiently trigger many-molecule reactions \cite{galego2017many} and alter photochemical processes \cite{campos2019resonant}. These applications have also spurred the development of fundamental theories on plexcitonic systems \cite{feist2018polaritonic,franke2019quantization}, bringing new vigor and vitality into the field of plasmonics.

In practice, achieving plexcitonic strong coupling with individual nanoantennas and arrays thereof demands a careful system design with an ultralow mode volume, extreme near-field enhancement, and an emitter whose transition dipole moment is precisely aligned with the maximum field direction. Among the various plasmonic nanocavity systems that have been investigated to date \cite{xiong2021room}, the so-called nanoparticle-on-mirror \cite{Ciracietal2012,Tserkezisetal2015NPoM,Kongsuwanetal2020QNM} (NPoM) configuration has emerged as one of the most successful, in which strong coupling at the single-molecule limit is enabled by the use of an extremely small ($<0.9$ nm) metal-insulator-metal gap, in conjunction with supramolecular guest-host chemistry to accommodate a single dye molecule \cite{chikkaraddy2016single,kongsuwan2018suppressed,Ojambatietal2019,Baumbergetal2019rev}. However, the use of such nanogap systems poses a number of technical challenges. First, individual dye molecules generally present small dipole moments, and aligning the molecule within the nanogap often relies on a probabilistic approach \cite{chikkaraddy2016single}. Second, such nanogaps cannot readily accommodate more sizeable quantum emitters or their ensembles, including quantum dots or aggregates of large dye molecules. Fortunately, two-dimensional (2D) materials, such as semiconducting transition-metal dichalcogenides (TMDCs) in monolayer form \cite{MakShan2016,Manzelietal2017}, have emerged as promising candidate quantum matter for plexcitonic strong coupling in plasmonic nanogaps, by virtue of their atomic-scale thickness ($\sim 3$\,\AA) together with the room-temperature stability and large dipole moments of their excitons. However, even if such ultrathin materials can be accommodated in a nanogap, their coupling to the cavity is still inefficient due to the misalignment of their in-plane excitonic dipole moments with respect to the direction of the dominant cavity field \cite{kleemann2017strong}. As a result, one may be compelled to compromise the field enhancement of the cavity by increasing the gap size \cite{kleemann2017strong,han2018rabi,hou2019manipulating}, such that an in-plane electric field component is allowed, or to harness the interlayer excitons supported by, for instance, heterostructures with type II band alignment, which often have only low oscillator strengths \cite{Fangetal2014PNAS,Meckbachetal2018,Torunetal2018}. Moreover, the NPoM and other contemporary cavity designs such as the cuboid Au@Ag nanorod \cite{liu2017strong} are usually fabricated via chemical assembly, which poses a significant challenge to produce devices on an industrial scale.
Notably, recent experiments on plexcitonic strong coupling have emerged based on nanostructures fabricated via top-down lithography \cite{inen2014plasmonic,santhosh2016vacuum,liu2016strong,kang2018strong,shan2019direct,gupta2021complex}. Such a technique allows for deterministic control of plasmonic nanosystems and may enable large-scale fabrication of plexcitonic chips, thus opening a route to practical applications of strong coupling. Towards the ultimate realization of plexcitonic quantum technologies and with a view to the potential of top-down fabrication approaches, we are motivated to explore plasmonic nanocavities with the following features: (1) hotspots that are open to excitonic matter; (2) plasmon-enhanced near-fields that are strongly aligned with excitonic dipoles; and (3) a design that is scalable and conducive to mass production.

In this work, we propose a novel concept based on a gold bowtie nanoantenna for plasmonic nanocavity design where plasmonic hotspots can be engineered via substrates. Here, we consider a bowtie placed on gold (Au), silicon (Si), and glass substrates, in comparison with a bare bowtie without substrate (\textit{i.e.}, suspended in air). Numerical simulations demonstrate that the hotspot in the gap region can be strongly enhanced and, more importantly, elevated to the upper surface of the bowtie by using Au and Si substrates, in stark contrast to the extensively studied bowtie-on-glass cavity. Such hotspots are more exposed and thus more readily accessible to quantum emitters. Our proposed scheme is particularly favorable for achieving a strong dipole coupling with 2D materials, since the dominant electric field is directed along the longitudinal axis of the bowtie (\textit{i.e.}, aligned with the in-plane excitonic dipoles in 2D materials). The studied plasmonic modes are also radiative with a good out-coupling efficiency to free space. Using a single bowtie placed on either a Au or Si substrate, we demonstrate strong coupling between their plasmonic modes and quantum matter comprising a single emitter or an ensemble thereof. The bowtie-on-Au and bowtie-on-Si cavities provide hotspots localized to the upper surface of the structure, and can potentially be fabricated via a top-down etching approach.

\section{II. Substrate-dependent Plasmonic Response of the Bowtie Cavity}

\begin{figure}[!ht]
\centering
\includegraphics[scale=0.42]{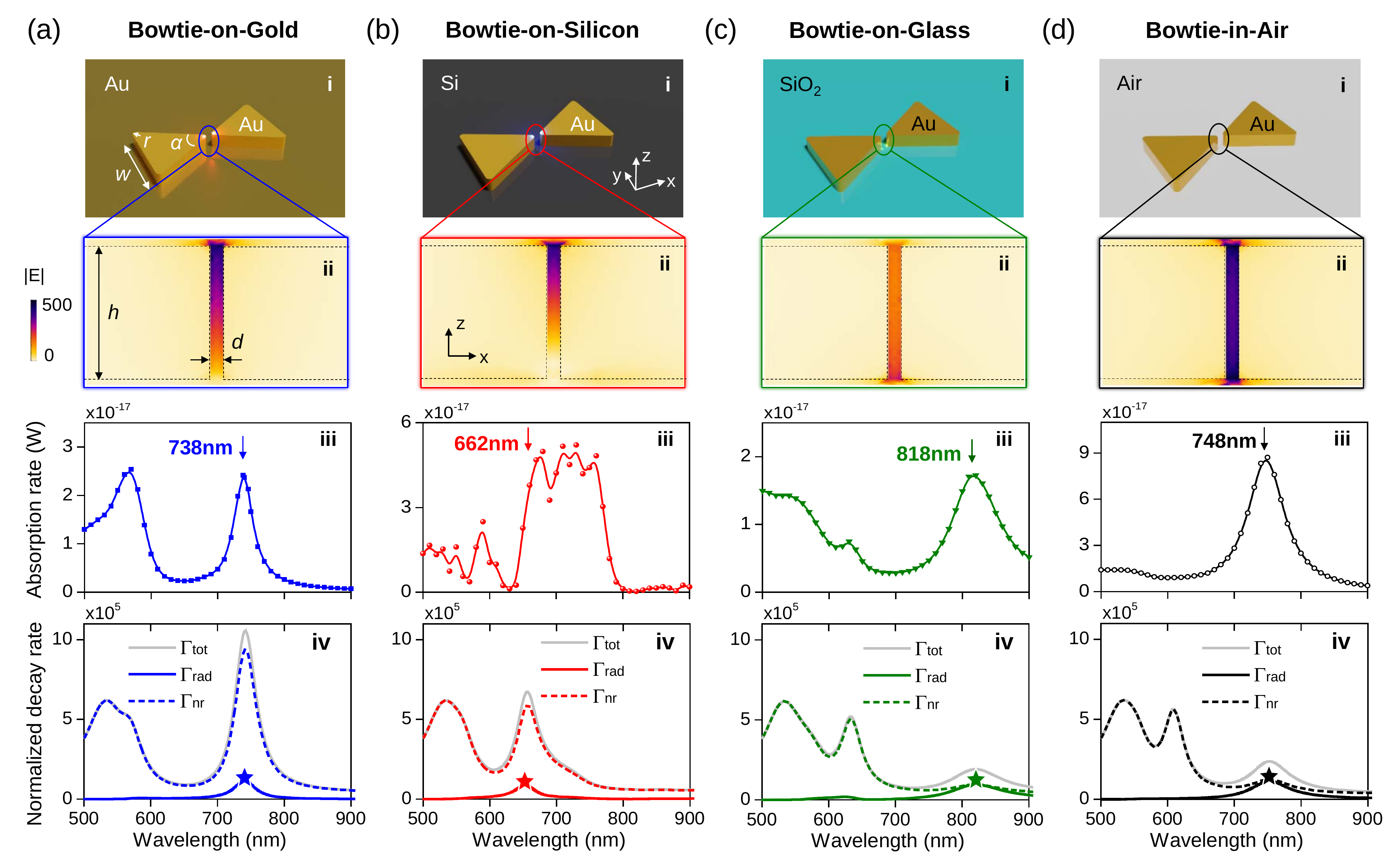}
\caption{
\textbf{Hotspot nanoengineering via substrates.} 
Gold nanobowties placed on different substrates: (a) gold, (b) silicon, and (c) glass; in comparison with (d) a bare gold nanobowtie in air.
(i) Schematics of the nanobowtie devices.
(ii) Spatial distributions of the normalized electric near-field within the nanobowtie gaps (side view, magnified) at their respective resonant wavelengths, with the geometric boundaries of each nanobowtie indicated by dashed black lines. 
(iii) Spectral absorption rates for the nanobowties, with arrows indicating the resonant wavelength for each configuration. Solid lines correspond to a spline interpolation of the data points in each case.
(iv) Spectral decay rates for radiative ($\Gamma_{\mathrm{rad}}$, colored solid curves) and non-radiative ($\Gamma_{\mathrm{nr}}$, colored dashed curves) channels, together with their total ($\Gamma_{\mathrm{tot}}$, gray curves), normalized with respect to the decay rate of a dipole source in free space. The bowtie geometry is identical throughout with $w=100$ nm, $h=30$ nm, $d=2$ nm, $\alpha=60^{\circ}$, and $r=5$ nm.
}
\label{fig1}
\end{figure}

To elucidate the substrate-dependent optical response of the Au bowtie nanocavity, we solve Maxwell's equations numerically by the finite-element method, as described in the Supporting Information (see Sec. S1). The Au bowtie is characterized by a set of geometrical parameters: bowtie width $w$, height $h$, gap distance $d$, apex angle $\alpha$, and radius of curvature $r$ at the corners, as illustrated in Fig. \ref{fig1}(a). In the following discussion, we fix the bowtie geometry with $w=100$ nm, $h=30$ nm, $d=2$ nm, $\alpha=60^{\circ}$, and $r=5$ nm, for simplicity. The detailed dependence of the optical response on the bowtie geometry can be found in the Supporting Information (see Sec. S2). The calculated spectral absorption rates for the Au bowtie are shown in Figs. \ref{fig1}(a)iii-\ref{fig1}(d)iii, with each exhibiting a multipeak structure that is sensitive to the choice of substrate. Note that, for the bowtie-on-Si cavity, there is a structured plateau feature around 700 nm, which has been discussed in the Supporting Information (see Sec. S1).
For each substrate and incident wavelength, we analyzed the electric near-field distribution around the nanobowtie, and found the maximum field enhancement at the wavelengths designated by the arrows. The corresponding normalized near-field profiles in the gap region are displayed in Figs. \ref{fig1}(a)ii-\ref{fig1}(d)ii. Note that, unless explicitly stated otherwise, the electric field data (both $|{\bf E}|$ and the individual field components) are normalized by the amplitude of the incident light field, $|\mathbf{E}_0| = 1$ V/m, and are thus unitless. It can be seen that the hotspots for the nanobowtie-on-Au and nanobowtie-on-Si cavities are lifted to the top of the gap, with the normalized $|{\bf E}|$ (\textit{i.e.}, field enhancement) for the Au substrate slightly larger. For the bowtie-on-glass cavity [Fig. \ref{fig1}(c)ii], the hotspot is located close to the substrate due to the higher refractive index of glass compared to the background air, while for the bare nanobowtie [Fig. \ref{fig1}(d)ii], the field enhancement is spatially uniform due to the symmetric environment. Compared to the glass substrate, the Au and Si substrates facilitate the elevation of the hotspot and, more importantly, enable a higher field enhancement of $\sim$ 1.6 times ($|{\bf E}|_{\mathrm{max}}\sim$ 482 for bowtie-on-Au, $|{\bf E}|_{\mathrm{max}}\sim$ 468 for bowtie-on-Si, and $|{\bf E}|_{\mathrm{max}}\sim$ 306 for bowtie-on-glass). Both features are extremely desirable in the practical pursuit of room-temperature, plexcitonic strong coupling.

We have also calculated the radiative ($\Gamma_{\mathrm{rad}}$), non-radiative ($\Gamma_{\mathrm{nr}}$), and total ($\Gamma_{\mathrm{tot}}$) decay rates for a classical dipole, aligned along the $x$-axis and located at the center of the gap along the bowtie upper surface. As shown in Figs. \ref{fig1}(a)iv-\ref{fig1}(d)iv, each configuration supports a radiative mode (highlighted with star symbols), whose wavelength is in good agreement with a peak in the corresponding absorption spectrum. This indicates that these modes not only possess an optimal field enhancement, but also couple efficiently to free space, thus allowing for far-field manipulations. Specifically, $\Gamma_{\mathrm{rad}}^{\mathrm{Au}}>\Gamma_{\mathrm{rad}}^{\mathrm{air}}>\Gamma_{\mathrm{rad}}^{\mathrm{glass}}>\Gamma_{\mathrm{rad}}^{\mathrm{Si}}$. 
Although the radiative yield $\Gamma_{\mathrm{rad}}^{\mathrm{Si}}$ is the lowest, it is still comparable ($\sim 10^5$) to $\Gamma_{\mathrm{rad}}^{\mathrm{Au}}$ for the bowtie-on-Au cavity.
With a Lorentzian fitting, the decay rates of the plasmon modes (half-width at half-maximum) are found to be $\kappa_\mathrm{c}^\mathrm{Au}=42.08$ meV for the bowtie-on-Au cavity and $\kappa_\mathrm{c}^\mathrm{Si}=57.58$ meV for the bowtie-on-Si cavity, respectively.
It should be noted that in the absorption spectra there are also other resonant peaks representing higher-order modes. 
As an example, we provide the details of the non-radiative mode at 570 nm for the bowtie-on-Au cavity in the Supporting Information (see Sec. S3).

\section{III. Physical Origins of the Hotspot Elevation}

\begin{figure}[!ht]
\centering
\includegraphics[scale=0.35]{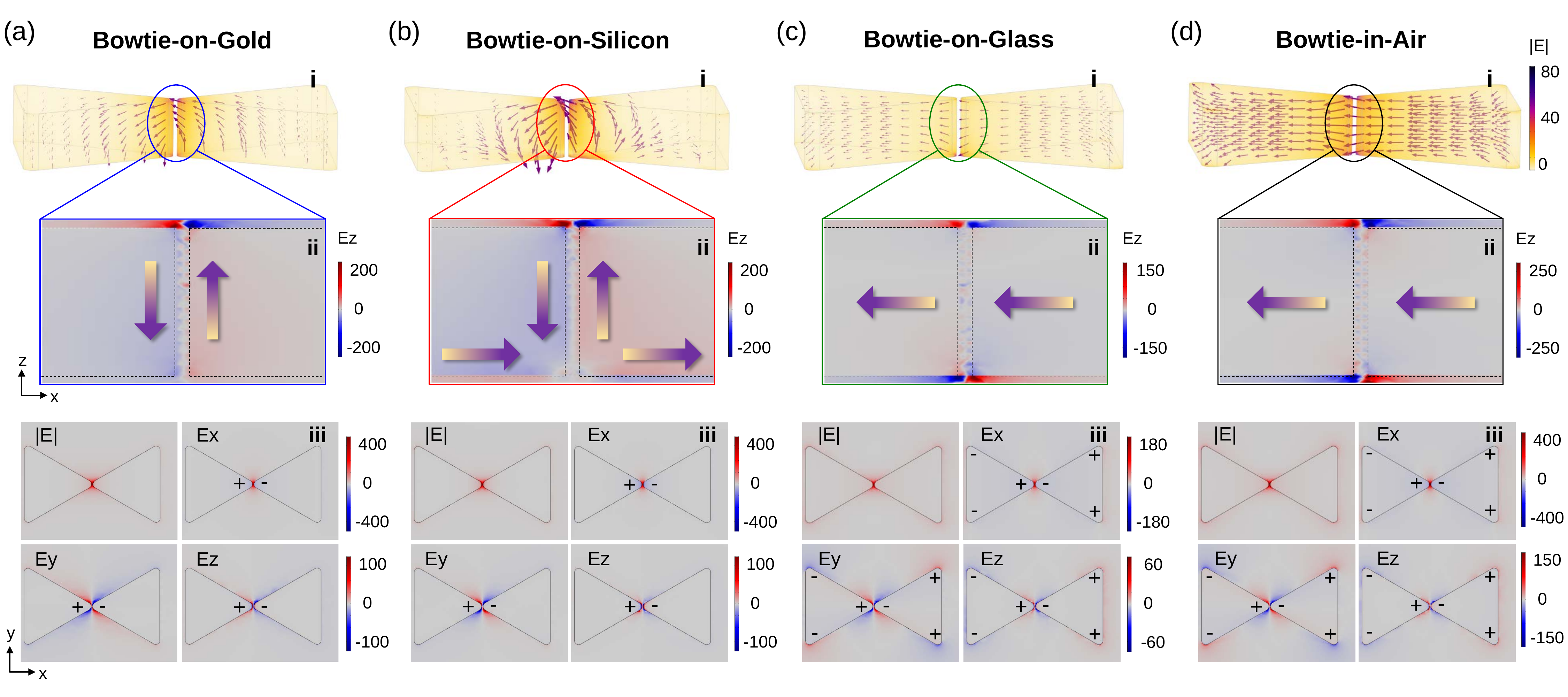}\\
\caption{
\textbf{Analysis of the substrate-dependent hotspot distributions.}
Details of the radiative modes supported by different substrates: (a) gold, (b) silicon, and (c) glass; in comparison with (d) a bare bowtie in air.
(i) Three-dimensional normalized electric near-field profiles of the bowties (color map), with superimposed vectorial electric field distributions (purple arrows) inside the devices. The arrow direction indicates that of the electric field, while its length is proportional to the magnitude of the electric field in a logarithmic scale.
(ii) The near-field distributions for the normalized $E_\mathrm{z}$ component in the gap region (side views), with bold, gradient arrows indicating the charge flow and dashed black lines indicating the geometric boundaries of the bowtie.
(iii) Near-field distributions for the normalized electric field magnitude and components at the upper surface of the bowties (top views), with superimposed charge distribution polarities. The field profiles show that $E_\mathrm{x}$ is the dominant component. The bowtie geometry is identical throughout with $w=100$ nm, $h=30$ nm, $d=2$ nm, $\alpha=60^{\circ}$, and $r=5$ nm.
}
\label{fig2}
\end{figure}

We performed in-depth studies of the corresponding radiative modes to better understand the physical origins of their substrate-dependent optical properties. Firstly, we extracted their three-dimensional normalized near-field profiles (color maps), as displayed in Figs. \ref{fig2}(a)i-\ref{fig2}(d)i. The field profiles show the locations of the hotspots: at the top of the gap for the bowtie-on-Au and bowtie-on-Si cavities, at the bottom of the gap for the bowtie-on-glass cavity, and uniform along the gap for the bare bowtie, which are consistent with the cross-sectional plots in Fig. \ref{fig1}. The normalized electric field distributions inside the bowtie structures are also presented as vectors (purple arrows), where the orientations and lengths of the arrows indicate the directions and magnitudes of the normalized electric field respectively. The charge flows (arrows) for the bowtie-on-Au and bowtie-on-Si cavities have a similar appearance, as do those for the bowtie-on-glass cavity and bare bowtie. For the former pair, the normalized electric field distributions suggest a dominantly vertical migration and accumulation of charge towards the upper tips of the nanoprisms, resulting in a stronger field enhancement (hotspot) at the upper surface of each structure. In contrast, for the latter pair, the charge motion has a largely horizontal character that leads to their accumulation along each of the opposing metal-dielectric interfaces, giving rise to a more spatially homogeneous field enhancement in the gap.

The same observation can also be made from the cross-sectional distributions for the $z$-component of the normalized electric field, $E_\mathrm{z}$, as shown in Figs. \ref{fig2}(a)ii-\ref{fig2}(d)ii. These unambiguously demonstrate that the electric field bears a significant vertical component for the bowtie-on-Au and bowtie-on-Si cavities (note the reddish and bluish coloration in the metal regions), implying a strongly vertical charge flow towards the upper vertices of the nanoprisms, as given by the bold, gradient arrows.
It should be emphasized that the charge flows are not identical for the Au and Si substrates [Figs. \ref{fig2}(a)i-\ref{fig2}(b)i]. For the bowtie-on-Si cavity, the internal electric field and thus the electronic charge motion bear some horizontal character away from the gap region, as indicated by an additional pair of bold, gradient arrows with such orientation in Fig. \ref{fig2}(b)ii. As for the bowtie-on-glass and bare bowtie structures [Figs. \ref{fig2}(c)ii-\ref{fig2}(d)ii], the normalized $E_\mathrm{z}$ component is much less significant in the metal regions (as indicated by the gray color), thus the bold, gradient arrows are all horizontal.

Additionally, Figs. \ref{fig2}(a)iii-\ref{fig2}(d)iii present the normalized near-field distributions at the upper surface of the bowtie structures for all electric field components. We note that the normalized field-component distributions in Fig. \ref{fig2}(c)iii for the glass substrate are qualitatively similar to those at the lower (contact) surface, but with a magnitude that is smaller by almost a factor of 2 ($|{\bf E}|_{\mathrm{max}}\sim$ 180 at the upper surface and $|{\bf E}|_{\mathrm{max}}\sim$ 306 at the lower surface). In the same figures, we also designate the polarity of the induced charge distributions according to the positive/negative signs of each component. Our findings suggest that the charge distributions for both the bowtie-on-Au and bowtie-on-Si cavities are strongly localized to the gap region, while some accumulation of charge along the exterior corners must also be present for the bowtie-on-glass cavity and bare bowtie. On this basis, we conclude that the radiative modes supported by the bowtie-on-glass cavity and bare bowtie are simply the conventional, bonding dimer plasmon modes that originate from the coupling between adjacent metal nanoprisms \cite{schmidt2014morphing}. In contrast, for the bowtie-on-Au and bowtie-on-Si cavities, the radiative modes bear strong contributions from the coupling between the substrate and nanoprisms, and are thus of a fundamentally different nature. Further discussion and characterization of the latter modes, including the origins of their elevated hotspots, can be found in the Supporting Information (see Sec. S4).
Finally, it is worth highlighting that in all cases, the $E_\mathrm{x}$ component is the dominant one, which is decisively favored for a strong dipole interaction with 2D materials, whose excitonic dipoles are typically oriented in-plane. However, among our investigated device configurations, the bowtie-on-Au and bowtie-on-Si cavities would enable the most efficient coupling compared to the bowtie-on-glass and bare bowtie, due to the $\sim$ 2.22 times larger field enhancement at the upper surface.

\section{IV. Antenna Mode for Strong Coupling}

\begin{figure}[!ht]
\centering
\includegraphics[scale=0.6]{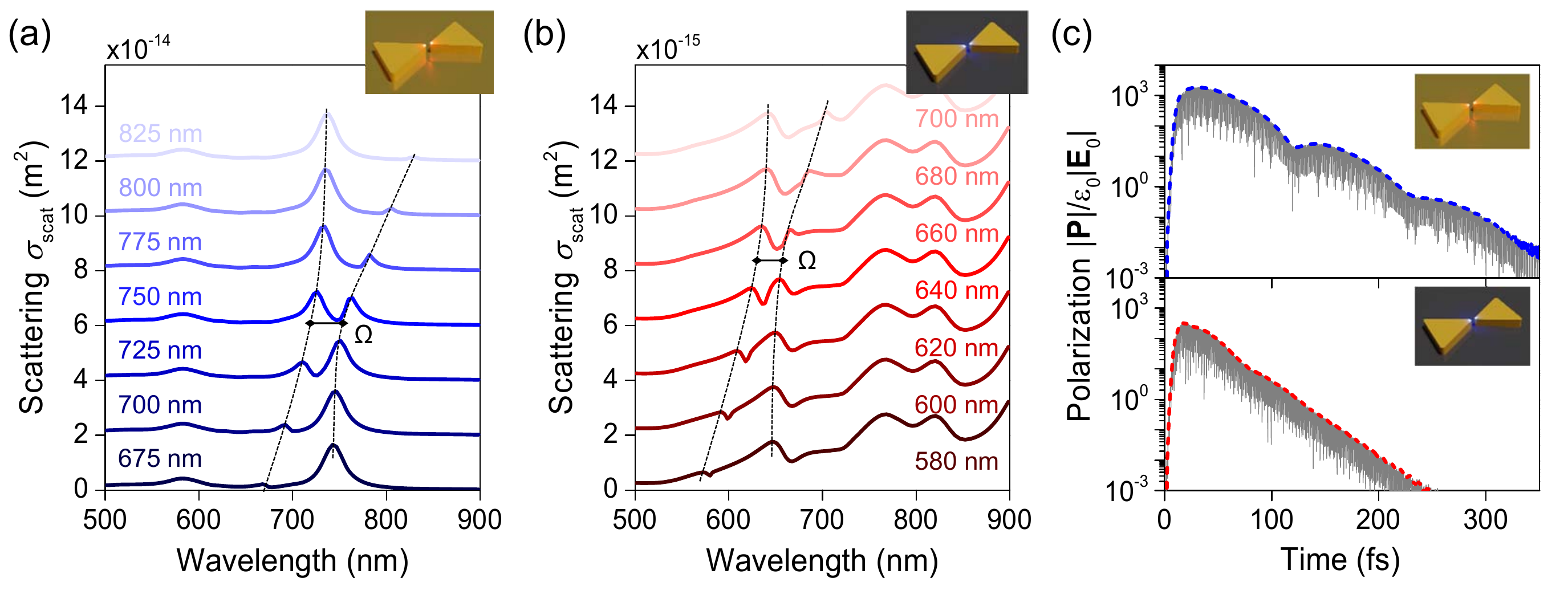}
\caption{
\textbf{Strong coupling of the bowtie-on-Au and bowtie-on-Si cavity antenna modes with a sheet of quantum emitters.} 
(a) The scattering spectra of the plexcitonic system for the bowtie-on-Au cavity (schematic in the inset) with the emitter resonance swept from 675 nm to 825 nm. The dashed lines trace the peaks in the scattering spectra, indicating the anticrossing behavior and Rabi splitting. 
(b) The scattering spectra of the plexcitonic system for the bowtie-on-Si cavity (schematic in the inset) with the emitter resonance swept from 580 nm to 700 nm. 
(c) The temporal dynamics of the normalized polarization $|\mathbf{P}|/\varepsilon_0|\mathbf{E}_0|$ for a single emitter placed in the nanogap hotspot of the bowtie-on-Au (top) and bowtie-on-Si (bottom) cavities. The dashed lines outline the envelope of $|\mathbf{P}|/\varepsilon_0|\mathbf{E}_0|$, where the oscillatory feature indicates a coherent exchange of energy between the cavity antenna mode and the quantum emitter.
The bowtie geometry is identical throughout with $w=100$ nm, $h=30$ nm, $d=2$ nm, $\alpha=60^{\circ}$, and $r=5$ nm, while the emitters have a fixed transition dipole moment of magnitude 5 D.
}
\label{fig3}
\end{figure}

In view of the dominantly in-plane polarized near-fields of the antenna mode and its potential for achieving strong light-matter interaction with 2D excitonic materials, such as semiconducting TMDCs and doped graphene, we next explore the performance of the bowtie-on-Au and bowtie-on-Si cavities in coupling active quantum media via their plasmonic hotspots. More specifically, we solve numerically the Maxwell-Bloch equations \cite{Boydbook}, adopted for nanoplasmonic materials and cavities  \cite{Hessetal2012active,Wuestner20141,Kongsuwanetal2019immunoassay}, for a thin sheet of uniformly distributed, two-level quantum emitters placed across the upper surface of each bowtie structure.
Details concerning the Maxwell-Bloch treatment and calculation parameters can be found in the Supporting Information (see Sec. S5). The emitters are identical and characterized by their transition wavelength $\lambda_{\mathrm{e}}$, decay rate $\Gamma_{\mathrm{e}}$ (half-width at half-maximum), and dipole moment $\mu$. We first fix the decay rate $\Gamma_{\mathrm{e}}=6.5$ meV (\textit{i.e.}, $10^{13}$ rad/s) and dipole moment $\mu=5$ D oriented along the $x$-axis, and study the scattering response of each coupled system as the emitter transition wavelength $\lambda_{\mathrm{e}}$ is swept across the cavity mode of interest. 

As shown in Figs. \ref{fig3}(a) and \ref{fig3}(b), the scattering spectra with swept $\lambda_{\mathrm{e}}$ exhibit a splitting feature for both the bowtie-on-Au and bowtie-on-Si cavities, as traced by the black dashed lines. The minimal splittings denoted $\Omega$ therein are obtained as $\Omega^\mathrm{Au}=82.78$ meV (at $\lambda_\mathrm{e}=747$ nm) and $\Omega^\mathrm{Si}=92.76$ meV (at $\lambda_\mathrm{e}=650$ nm).
Although such splitting is a well-known manifestation of strong coupling in the single- or many-emitter regimes, it could also originate from a coherent Fano interference effect under intermediate coupling conditions \cite{Lengetal2018Fano}. To provide firm evidence that our bowtie-on-Au and bowtie-on-Si cavities are capable of supporting light-matter strong coupling, we also investigate the interaction of the antenna mode with a single, two-level emitter positioned in the elevated hotspot of each cavity and subject to identical excitation conditions. The corresponding scattering spectra in the single-emitter regime are shown in Fig. S6 of the Supporting Information, where a dual-peak structure with an anticrossing behaviour is observed as the cavity resonance and emitter transition are progressively detuned.
To gain a deeper insight, we further examine the temporal dynamics of the normalized polarization for the emitter in each coupled system. As shown in Fig. \ref{fig3}(c), the polarization fluctuations in each case have a period of $\sim1.2$ fs, corresponding to the transition frequency of the emitter, and conform to an exponentially decaying envelope (dashed lines).
Crucially, this envelope is accompanied by an oscillatory modulation, which signifies a coherent energy cycling characteristic of plexciton formation in each system. For the bowtie-on-Au cavity, the oscillation period is $\sim115$ fs, corresponding to an energy exchange rate of 8.69 THz. In contrast, the oscillation is less pronounced for the bowtie-on-Si cavity, where only one round of energy exchange is observed. This is attributable to the multiplicity of resonant modes [particularly at longer wavelengths, see Fig. \ref{fig3}(b)] supported by the bowtie-on-Si cavity, which provide additional channels by which energy dissipates from the cavity antenna mode. As energy is transferred from the emitter to the antenna mode, it may quickly dissipate via other resonant modes that are spectrally or spatially overlapping with the antenna mode. The presence of Rabi oscillations constitutes an unambiguous signature of the plexcitonic regime in the case of a single quantum emitter, and we can thus identify the dual-peak features in Figs. S6(a) and S6(b) as polaritonic splittings. Invoking a phenomenological, dissipative Jaynes-Cummmings treatment \cite{xu2018quantum,Baranovetal2018review}, the latter can be related to the single-emitter coupling strength $g_{0}$ via $\Omega=\sqrt{4g^2_0-(\Gamma_\mathrm{e}-\kappa_\mathrm{c})^2}$, where $\Gamma_\mathrm{e}$ is the emitter decay rate defined above and $\kappa_\mathrm{c}$ is the decay rate of the cavity mode. The single-emitter coupling strengths are thereby deduced as $g^\mathrm{Au}_0=31.28$ meV for the bowtie-on-Au cavity, and $g^\mathrm{Si}_0=33.81$ meV for the bowtie-on-Si cavity, respectively (see Sec. S6). 

The ability of these plasmonic cavities to support strong coupling in the single-emitter regime suggests that the spectral splittings observed for an ensemble of many such emitters in Figs. \ref{fig3}(a) and \ref{fig3}(b) could also originate from a strong coupling behaviour, mediated by one or perhaps a collection of emitters interacting with the same antenna mode of each bowtie-on-substrate cavity.
In the absence of intrinsic cavity or emitter dissipative effects, the Tavis-Cummings model \cite{TavisCummings1967} yields the well-known prediction that the separation of the dressed states in such a scenario is given by $\Omega = 2g_\mathrm{eff}$, with an effective coupling strength $g_\mathrm{eff}=\sqrt{N}g_0$ enhanced relative to the single-emitter one $g_0$ according to the number $N$ of quantum emitters that collectively participate in the strong coupling. Given the complexity of treating dissipation in the many-emitter case, we shall adopt these simple relations in the present analysis to provide effective estimates for the strength and collectivity of light-matter interaction, albeit retaining the renormalizing effect of decay in the calculation of $g_0$ (as above).
From the splittings observed in Figs. \ref{fig3}(a) and \ref{fig3}(b), the effective coupling strengths are deduced as $g^\mathrm{Au}_\mathrm{eff}=41.39$ meV and $g^\mathrm{Si}_\mathrm{eff}=46.38$ meV, and the effective number of emitters can be estimated as $N^\mathrm{Au}\approx 2$ and $N^\mathrm{Si}\approx 2$.

According to the above analysis, the bowtie-on-Au and bowtie-on-Si cavities result in a similar coupling strength with the same quantum emitters, albeit at different emitter transition wavelengths $\lambda_\mathrm{e}$. They can therefore be used to achieve strong light-matter interaction in different spectral ranges and match with different quantum emitters. From an experimental perspective, the evaporation of a Au film in forming the substrate may lead to a rough surface with typical roughness values in the range 1.6 -- 4.0 nm, depending on the evaporation conditions \cite{Ajit2006Gold}. Such a high surface roughness may distort the designed plasmonic cavity resonance and impose stringent requirements on the nanofabrication process to realize sub-10-nm gaps in a uniform manner. On the other hand, the commercially available Si surface is usually atomically flat \cite{Dong2015Gap5nm} and thus the fabrication of dimers with $\sim 5$ nm gaps is feasible \cite{Dong2019Gap5nm}. It is also worth highlighting that the scattering cross-section of the bowtie-on-Si cavity is typically an order of magnitude smaller than that of the bowtie-on-Au cavity in the investigated spectral range. This is due to a (rather non-trivial) distribution of spectral weight among multiple resonant modes of the bowtie-on-Si cavity. Nevertheless, at longer wavelengths (1000 nm, beyond the range considered here), the scattering spectrum of this system is dominated by another mode with a peak cross-section $\sim 4.7$ times higher than that of the antenna mode. However, this mode displays a near-field gap hotspot located closer to the substrate, similar to the conventional bonding dimer plasmon mode of a Au bowtie device, and is therefore not of interest in the present study.


\section{V. Engineering of the strong coupling}

\begin{figure}[!ht]
\centering
\includegraphics[scale=0.55]{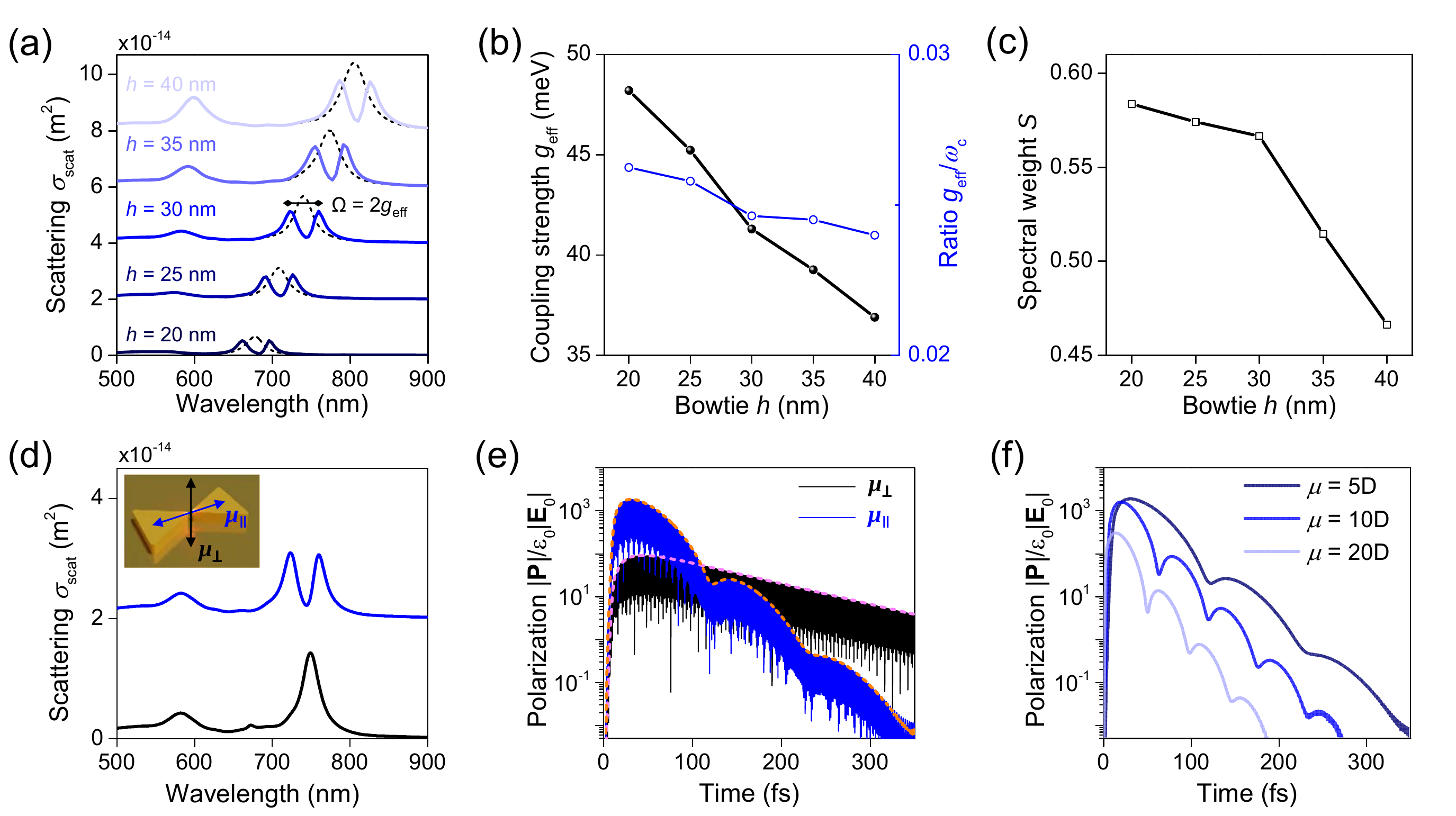}
\caption{
\textbf{Engineering of plexcitonic strong coupling with the bowtie-on-Au cavity.} 
(a) The scattering spectra of the plexcitonic system with varying bowtie height $h$. In each case, the emitter wavelength $\lambda_\mathrm{e}$ is tuned accordingly to achieve the minimal (Rabi) splitting. Dashed lines indicate the scattering spectra of the corresponding, bare bowtie-on-Au cavities.
(b) The effective coupling strength $g_\mathrm{eff}$ (black dots) extracted from (a) and the normalized coupling strength $g_\mathrm{eff}/\omega_\mathrm{c}$ (blue circles) as functions of $h$, where $\omega_\mathrm{c}$ is the frequency of the antenna mode. 
(c) Normalized spectral weight of the antenna mode as a function of $h$, calculated in accordance with Eq. (\ref{spectralweight}).
(d) The scattering spectra for emitters with in-plane (blue line) and out-of-plane (black line) transition dipole moments. Inset: the respective dipole orientations.
(e) The temporal dynamics of the normalized polarization $|\mathbf{P}|/\varepsilon_0|\mathbf{E}_0|$ for a single emitter at the bowtie tip with a fixed magnitude of the transition dipole moment, $\mu = 5$ D, but different orientations. Dashed lines indicate the envelope of the polarization dynamics in each case. For $\mu_\parallel$, the envelope bears an oscillatory modulation in addition to its exponential decay.   
(f) The temporal dynamics of the normalized polarization envelope for a single emitter at the bowtie tip with different transition dipole moments $\mu$ but fixed, in-plane orientation.
Unless stated otherwise, the reference bowtie geometry is identical throughout with $w=100$ nm, $h=30$ nm, $d=2$ nm, $\alpha=60^{\circ}$, and $r=5$ nm.
}
\label{fig4}
\end{figure}

Finally, we investigate the effect of the bowtie geometry and emitter characteristics on the strong coupling behaviour, and explore the possibility of engineering the light-matter coupling via these parameters. As shown in Fig. \ref{fig3}, the bowtie-on-Au cavity presents a stronger scattering signal and more clearly resolved resonant modes; we therefore focus on the bowtie-on-Au cavity in this section. Nevertheless, the engineering principles that will be discussed in the following are also applicable to the bowtie-on-Si cavity. 

Among the geometrical parameters of the bowtie, the height $h$, and apex angle $\alpha$ can be well controlled in practical lithographic device fabrication, and offer flexibility to engineer the strong coupling via the sensitive dependence of the plasmonic field enhancement and the resonant wavelength on these properties. In Fig. \ref{fig4}(a), the scattering spectrum of the bowtie-on-Au cavity coupled with a layer of single quantum emitters is plotted for different bowtie heights $h$. As $h$ increases, the global scattering signal becomes stronger, and the antenna mode resonance (black dashed lines) is also red-shifted. Accordingly, for different $h$ values, the emitter transition wavelength $\lambda_\mathrm{e}$ is also tuned to match with the antenna mode, thereby maintaining a Rabi splitting for the plexcitonic system (blue solid lines), denoted $\Omega$ and indicated in the figure. The effective coupling strength can then be extracted according to $\Omega=2g_\mathrm{eff}$.
Note that the mode at $\sim570$ nm is only weakly coupled with the quantum emitters, so that the  spectra in this region are barely changed (note the overlapping blue and black lines). Furthermore, this mode is less sensitive to variations in $h$, since it partially originates from the coupling between the adjacent nanoprisms of the bowtie (see Fig. S4 in the Supporting Information).

Figure \ref{fig4}(b) shows the extracted coupling strength $g_\mathrm{eff}$ as a function of the bowtie height (black line), which appears to decrease in a monotonic fashion as $h$ increases. Even if the coupling strength is normalized to the cavity resonance (blue line), $g_\mathrm{eff}/\omega_\mathrm{c}$ still exhibits a decreasing slope. This dependence is contrary to that of the local field enhancement $|{\bf E}|$ on $h$, which becomes stronger as $h$ increases (see Fig. S3 in the Supporting Information).
To better understand this behaviour, we focus on the optical response of the bare cavity, and examine in more detail the evolution of the plasmonic antenna mode with bowtie height. Specifically, we define a normalized measure for the spectral weight of each cavity mode in the form
\begin{equation}\label{spectralweight}
    S=\frac{\int_{\mathrm{mode}} \sigma_{\mathrm{scat}}(\lambda) \mathrm{d}\lambda}{\int \sigma_{\mathrm{scat}}(\lambda) \mathrm{d}\lambda},
\end{equation}
where $\sigma_{\mathrm{scat}}(\lambda)$ is the scattering cross-section shown in Fig. \ref{fig4}(a) (black dashed lines), and the integration in the numerator is performed over the spectral range pertaining to the mode in question, while that in the denominator is performed over the entire spectrum. The quantity $S$ is plotted as a function of $h$ in Fig. \ref{fig4}(c), and exhibits a decreasing trend similar to the coupling strength itself. Our results imply that the higher-order mode around 570 nm accumulates spectral weight relative to the antenna mode as $h$ increases, and is thus excited more efficiently. This effect appears to dominate the larger field enhancements that would otherwise give rise to a larger splitting for higher $h$ values, producing the observed decrease of the coupling strength with bowtie height.
On the other hand, this higher-order mode also couples to the layer of quantum emitters via its spatially overlapping hotspots (see Fig. S4 in the Supporting Information). As $h$ increases however, the red shift in the antenna mode requires the emitter transition wavelength to be increased also for spectral matching purposes, which in turn increases its detuning relative to the higher-order mode and compromises their interaction strength. Besides the bowtie height, the apex angle $\alpha$ can also be used to fine-tune the plasmonic resonance while maintaining the plexcitonic coupling. Detailed discussions can be found in the Supporting Information (see Figs. S3 \& S7).

The properties of the quantum emitters provide an additional flexibility to engineer plexcitonic coupling, particularly through the magnitude and orientation of their transition dipole moments. We denote in-plane and out-of-plane orientations of the emitter dipole moment by $\mu_\parallel$ and $\mu_\perp$, respectively, and show the corresponding scattering spectra of the coupled system in Fig. \ref{fig4}(d). A prominent spectral splitting is incurred for the emitters with $\mu_\parallel$ (blue line), while the spectrum for $\mu_\perp$ (black line) almost resembles the optical response of the bare cavity [dashed line in Fig. \ref{fig4}(a)]. These results unambiguously demonstrate an efficient plexcitonic interaction due to the favourable alignment between the dominantly in-plane plasmonic cavity field [Fig. \ref{fig2}(a)iii] and the emitter dipole $\mu_\parallel$, and further suggest that the bowtie antenna mode is appealing for coupling 2D excitonic materials, whose transition dipoles are largely in-plane.

Having predicted the feasibility of a collective light-matter strong coupling using the bowtie-on-Au cavity, we now explore the fundamentally and technologically interesting regime of strong coupling in the single-emitter limit \cite{xiong2021room}. We position a single, two-level quantum emitter in the gap hotspot of the bowtie-on-Au cavity and explore its temporal dynamics for different magnitudes and orientations of the transition dipole moment. A detailed description of the simulation methodology and numerical parameters can be found in the Supporting Information (see Sec. S5). Figure \ref{fig4}(e) presents the normalized polarization of the emitter as a function of time for two orthogonal dipole orientations. Clearly, the polarization envelope for $\mu_\parallel$ shows a faster decay than that for $\mu_\perp$, implying a stronger coupling with the cavity. More importantly, the oscillatory modulation (period of $\sim115$ fs) in the envelope for $\mu_\parallel$ signifies a coherent energy cycling within the plexcitonic system and further proves the dominant in-plane nature of the field for the antenna mode. 
When the dipole moment is increased beyond 5 D, as shown in Fig. \ref{fig4}(f), the energy exchange occurs at a higher rate with more oscillations in the investigated temporal interval, together with an overall faster exponential decay. We note that the values of $\mu$ considered here span those of several real quantum emitters including large dye molecules and colloidal quantum dots, while even higher excitonic transition moments have been observed in monolayer TMDCs like WS$_{2}$ \cite{xiong2021room}. As such, our proposed bowtie-on-Au cavity appears to hold excellent potential as a platform for plasmonic cavity quantum electrodynamics and plexcitonic quantum devices, even in the single-emitter limit.

\section{VI. Conclusions and Outlook}

To conclude, we propose a three-dimensional engineering of plasmonic cavity hotspots at the nanometer scale via substrate selection. By placing a Au nanobowtie on Au and Si substrates, we observe an elevation of the hotspots to the exterior/upper surface of the nanobowtie, thus rendering them more readily accessible for quantum emitters to achieve plasmon-enhanced light-matter interaction and a myriad of its pertinent practical applications. This hotspot elevation originates from the antenna mode that involves a coupling between the Au bowtie structure and substrate, and is fundamentally different from the widely explored bonding dimer plasmon mode supported by the bare nanobowtie or nanobowtie-on-glass cavity. Another feature of the antenna mode is the dominant in-plane electric field, which is especially favorable for interaction with excitonic 2D materials compared to, \textit{e.g.}, the popular NPoM cavity. Using this antenna mode, we demonstrated that strong coupling can be achieved simply by placing a layer of quantum emitters on top of the nanobowtie, unveiling the prospect of observing the plexcitonic regime in atomically thin materials such as monolayer TMDCs and doped graphene.
Our system design is conducive to a high degree of engineering flexibility for coupling active quantum media, where both the bowtie geometry and choice of substrate enable spatiospectral tunability of the plasmonic excitations. Furthermore, our design allows top-down fabrication, enabling scaling-up and mass production of plexcitonic chips, paving the way to room-temperature plexcitonic quantum technologies.

Aside from the realization of strong coupling with 2D materials, we also envisage that the investigated antenna mode will be harnessable in other contemporary applications. For example, leveraging the large spatial field gradients in the cavity, the antenna mode might be exploited for field-enhanced single-molecule spectroscopy. With the advent of precision methods for deterministic orientation of single molecules in nanogaps \cite{chikkaraddy2016single}, mechanisms for controlled hotspot distribution allow different regions of an individual molecule to experience different local electromagnetic environments, which could, in turn, be exploited for temporally and/or spatially resolved spectroscopic characterization of different excitonic states. Recent research has also shown the possibility to create strong coupling with living organisms for the study of biological processes, \textit{e.g.}, photosynthesis or bacterial growth \cite{coles2014strong}, since many electronic transitions in biologically relevant molecules occur in the ultraviolet and visible ranges. Multidimensional hotspot manipulation could offer greater flexibility for novel spectroscopic opportunities here as well.

The underlying principles of hotspot/substrate nanoengineering are general and can be readily applied to other metals and plasmonic nanoparticles. By using a different metal or changing the geometry of the antenna, the plasmonic resonance can be tuned to other spectral ranges for a variety of potential applications. For example, in the ultraviolet range, aluminum-based plasmonic cavities \cite{knight2014aluminum} may be designed to enhance the efficiency of photocatalysis \cite{ghori2018role} or photovoltaic energy conversion \cite{li2017harvesting,dubey2020aluminum}, enabling low-cost, CMOS-compatible plasmonic devices in an environmentally-friendly and sustainable economy. By coupling the antenna mode with 2D materials, powerful optoelectronic devices can be designed for mid-to-far infrared range functionality \cite{yu2018atomically,yu2018narrow,lukman2020high}, and may benefit a broad spectrum of applications, such as environmental monitoring, health care diagnostics and non-destructive inspection.

Substrate engineering could also facilitate novel opportunities for tailoring the functionality of diffractive arrays and metasurfaces. In particular, the high quality factor, spatially delocalized character and unique dispersion properties of surface lattice resonances \cite{liu2016strong,liu2019observation,li2019large} supported by nanoantenna arrays render them as interesting candidates for polaritonic device technologies, but the additional design scope conferred by the substrate material has not yet been thoroughly explored. Recently, lattices of localized polariton condensates have shown the capability of forming topologically protected states \cite{st2017lasing,klembt2018exciton} and simulating many-body Hamiltonians \cite{berloff2017realizing}. The complexity of such lattices also makes them superior in neuromorphic computing, which can be faster and more power-efficient than the von Neumann architecture \cite{ballarini2020polaritonic}. Our proposed material-based engineering of plasmonic nanocavity modes should create new design flexibility for emerging plexcitonic lattices, leveraging the potential of strongly interacting light-matter systems for fully parallel and ultrafast operations in novel artificial intelligence systems.

\begin{acknowledgement}
The Institute of High Performance Computing (IHPC) acknowledges financial support from the A*STAR Strategic Program (No. C210917001).
D.C. and O.H. gratefully acknowledge funding from Science Foundation Ireland via Grant No. 18/RP/6236.
L.W. gratefully acknowledges the Start-Up Research Grant from Singapore University of Technology and Design via Grant No. SRG SMT 2021 169.
The computational work reported in this article relied on support and infrastructure provided by the Trinity Centre for High Performance Computing, with funding from the European Research Council, Science Foundation Ireland and the Higher Education Authority, through its PRTLI program.
\end{acknowledgement}


\begin{suppinfo}

The following files are available free of charge at https://pubs.acs.org/doi/xxxx.
\\
Numerical Models for Near-field Studies;
Tunability of the Optical Response via the Bowtie Geometry;
Higher-order Non-radiative Mode of the Bowtie-on-Au Cavity;
Different Mechanisms for Hotspot Formation using Au and Si Substrates; 
Two-level Maxwell-Bloch Simulations;
Scattering Spectrum in the Single-emitter Strong Coupling Regime;
Dependence of the Plexcitonic Rabi Splitting on the Bowtie Apex Angle;
Figures S1-S7 (PDF).

\end{suppinfo}


\bibliography{reference}

\end{document}


\maketitle

\renewcommand{\thefigure}{S\arabic{figure}}
\renewcommand{\thetable}{S\arabic{table}}
\renewcommand{\theequation}{S\arabic{equation}}


\clearpage
\section*{\large S1. Numerical Models for Near-field Studies}

\subsection{\small Details}

Our numerical simulations for the substrate-dependent electromagnetic response of each nanobowtie device reported in Sections II and III of the main text were performed using COMSOL Multiphysics \cite{COMSOLwebsite}, and based on two distinct models for comparison and validation purposes. We designate these as Model I and Model II henceforth. Model I is a standard Mie scattering description \cite{mie}. We assume that an $x$-polarized, electromagnetic plane wave is incident in the $z$-direction (i.e., normally) on the bowtie, which is placed on a large substrate (Au, Si, glass) or in air. The wavelength-dependent refractive indices of Au and Si are taken from the literature \cite{johnson1972optical,schinke2015uncertainty}, while the refractive indices of glass and air are fixed at $n=1.45$ and $n=1$ respectively. The substrate has an area of $8w\times 8w$ (with $w$ the bowtie width), and a finite thickness of $t=100$ nm. The bowtie together with the substrate form the scatterer, and our simulations entail the calculation of the electromagnetic near-fields, the total scattered fields and experimentally measurable quantities like the power absorption spectrum and scattering cross-section. In contrast, Model II treats the bowtie as the scatterer and considers the substrate to occupy the entire $z<0$ half-space (i.e., an infinite substrate) \cite{scatterer}. We again assume normal incidence in the $z$-direction and x-polarization as for Model I, but now impose periodic boundary conditions in the $x$ and $y$ directions to simulate a substrate and optical source of infinite lateral extent. In the main text, the results presented in Figs. 1-2 are obtained from Model I.

It is worth noting that for few-nanometer gap sizes (down to 2 nm in this study), quantum effects such as electron spill-out and tunneling across the bowtie junction are expected to play only a minor role, such that a purely classical treatment of the plasmonic modes supported by the bowtie cavity should be justified. Moreover, the spatial non-locality inherent in the electromagnetic response of the system is expected to incur only a limited correction to the predicted field enhancements, particularly at the prism apexes, and should not qualitatively alter the conclusions of our analysis.

\begin{figure}[!ht]
\centering
\includegraphics[scale=0.6]{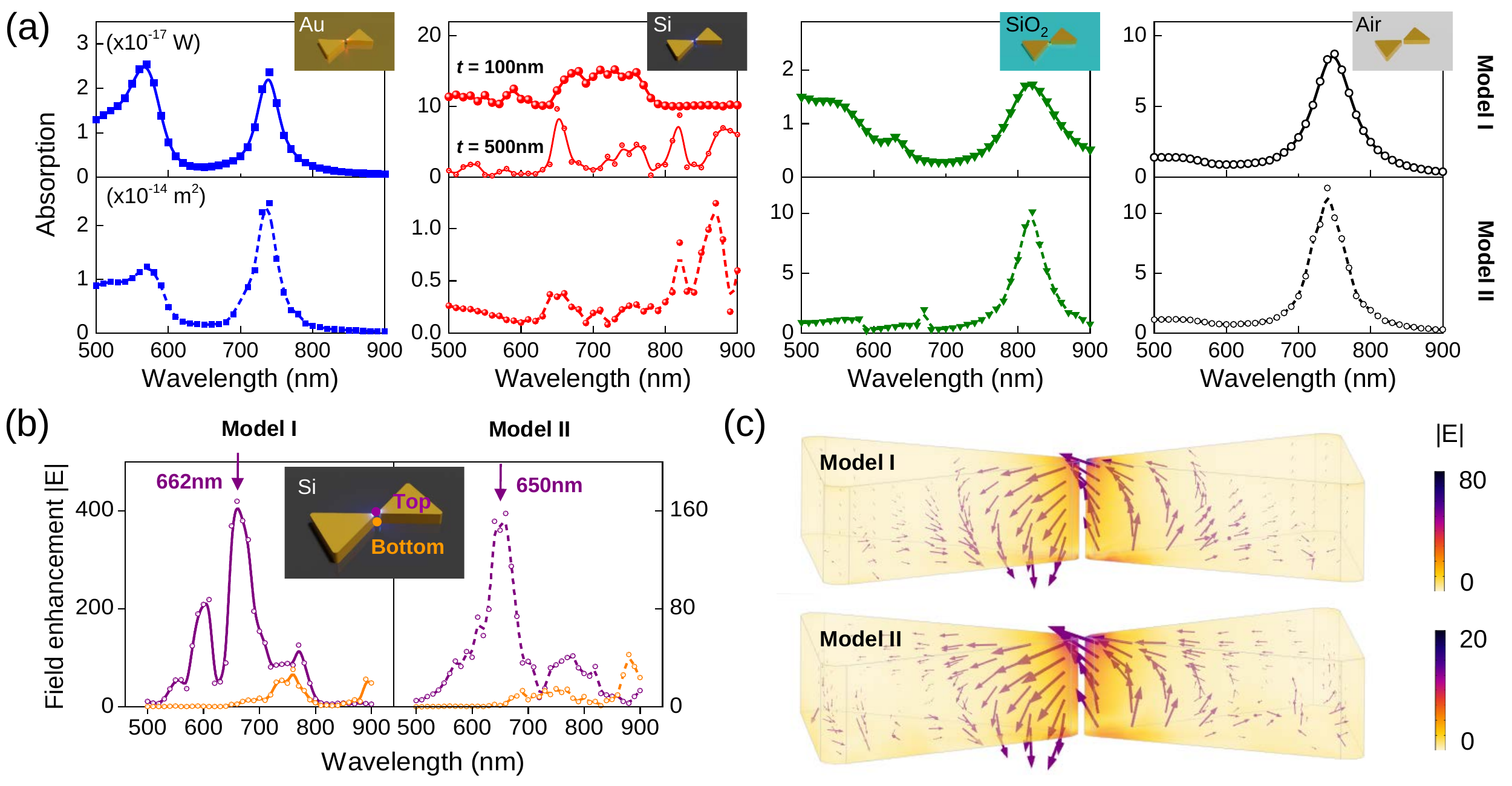}\\
\caption{
\textbf{Comparison between Mie scattering (Model I) and scatterer on a substrate (Model II) descriptions.}
(a) Absorption spectra of the nanobowtie placed on a Au substrate, a Si substrate, a glass substrate or suspended in air, obtained from Model I (top panel) and Model II (bottom panel). In the second column, absorption spectra are shown for Si-substrate thicknesses of $t=100$ nm and 500 nm, with the former offset by $10\times10^{-17}$ W. In all other cases, the substrate thickness in Model I is fixed at $t = 100$ nm. (b) The field enhancements $|E|$ at the top (purple lines) and bottom (orange lines) of the gap center, extracted from Model I (left panel) and Model II (right panel). In both models, the hotspots are located at the upper surface of the nanobowtie at a wavelength of $\sim 655$ nm. (c) The three-dimensional, normalized electric near-field profiles of the nanobowtie (color map), with superimposed vectorial electric field distributions (or charge flows) inside the device (purple arrows). The arrow direction indicates that of the electric field, while the arrow length is proportional to its magnitude in a logarithmic scale. Note that in the case of Model II, the arrow lengths are scaled by a factor of 3 compared to those for Model I. The geometrical parameters of the nanobowtie are $w=100$ nm, $h=30$ nm, $d=2$ nm, $\alpha=60^{\circ}$, and $r=5$ nm. 
}
\label{figs1}
\end{figure}

The absorption spectra of the bowtie are compared for both models in Fig. \ref{figs1}(a). Note that for Model I, we record the power absorption in the bowtie (units of W), while for Model II we normalize it by the absorption cross-section (units of m$^2$) given the infinite irradiation area. Overall, the two models show good qualitative agreement with each other except for the case of a Si substrate, which we analyze further in the following.

\subsection{\small Discrepancy for Si Substrate}

As shown in the second column of Fig. \ref{figs1}(a), Models I and II give rather different absorption spectra for the bowtie-on-Si cavity when a substrate thickness of $t=100$ nm is used in the former. In that case, the spectrum shows a collection of strongly overlapping features and lacks clear mode definition. The poorer agreement between Models I and II for the Si substrate compared to the other designs may be due to the higher refractive index of Si itself, such that the incident light experiences the finite substrate thickness and behaves differently. Upon increasing the thickness of the substrate from $t=100$ nm to 500 nm, we have observed a smooth evolution in the spectrum giving rise to improved mode resolution. In particular, the spectrum for $t=500$ nm allows much clearer resonance identification, especially with respect to the antenna mode around 662 nm, and shows improved resemblance to that found in Model II (for an infinite substrate).

In addition, we have analyzed the electric near-fields for each wavelength, and found that the hotspots at the upper surface of the bowtie as well as their formation mechanisms are well captured by Model I, even when using a substrate thickness of $t=100$ nm. Figure \ref{figs1}(b) presents the field enhancements $|E|$ at both the top and bottom of the gap center as a function of wavelength. We observe that the hotspots with maximal field enhancement are located at the upper surface in both models, and more importantly, at similar wavelengths (662 nm in Model I and 650 nm in Model II). Figure \ref{figs1}(c) further shows the electric near-field profiles at these wavelengths with superimposed charge flows, which unambiguously suggest that the modes attaining the aforementioned hotspots are the same in both models, irrespective of the smaller maximum value of $|E|$ in Model II. Based on these findings, we conclude that the Mie scattering model (Model I) is valid for our investigation of the near-field characteristics of the plasmonic modes of interest in the main text. In fact, Model I has been used previously for analyzing the plasmonic properties of gold nanocuboids \cite{huang2013synthesis}, and showed good agreement with experimental measurements.

\clearpage
\section*{\large S2. Tunability of the Optical Response via the Bowtie Geometry}

\begin{figure}[!ht]
\centering
\includegraphics[scale=0.7]{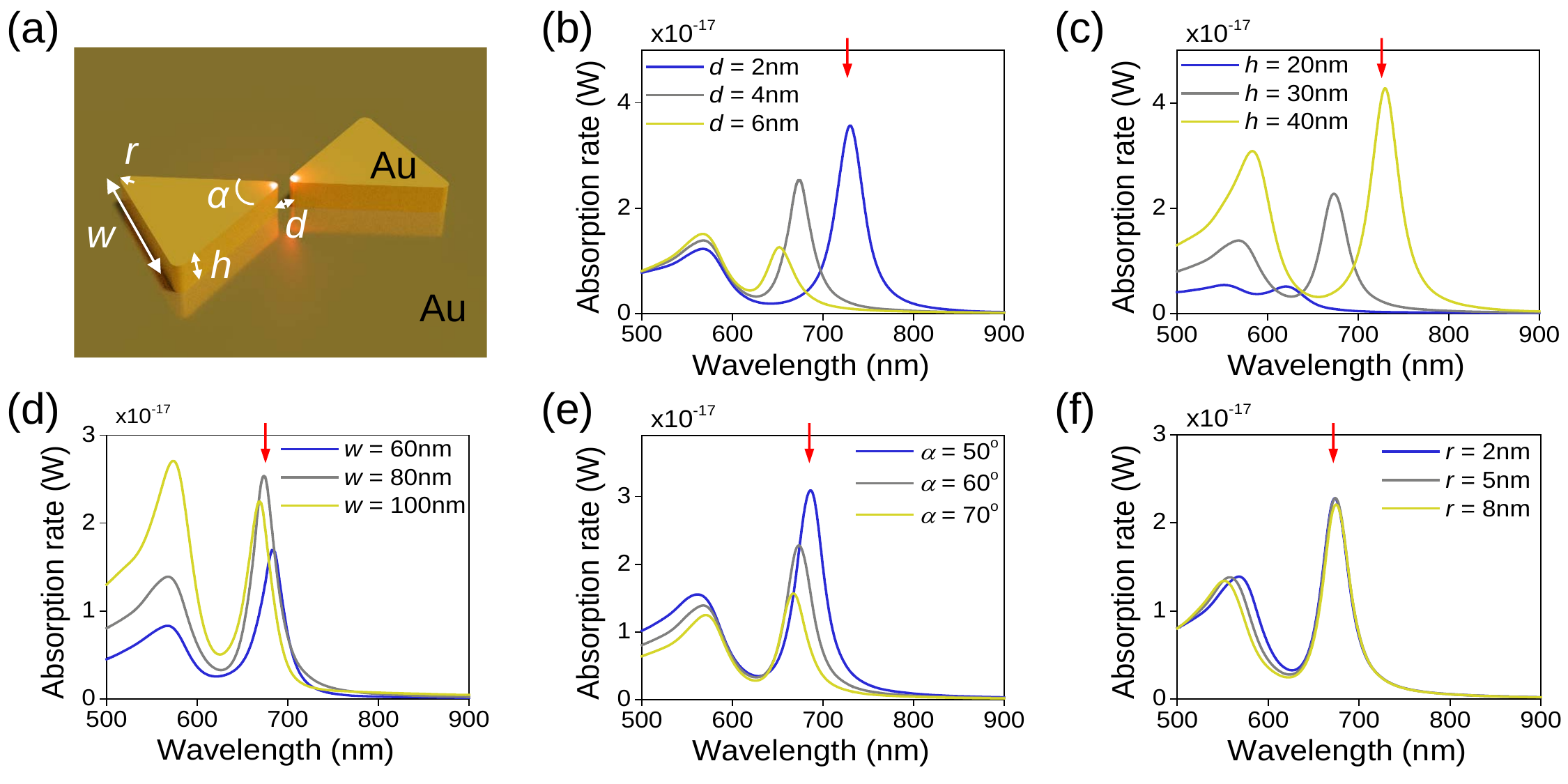}\\
\caption{
\textbf{Dependence of the bowtie optical response on geometry.}
(a) Schematic of the bowtie-on-Au cavity, with relevant geometrical parameters as indicated. (b-f) Variation of the bowtie absorption spectrum as a function of (b) gap size $d$, (c) height $h$, (d) width $w$, (e) apex angle $\alpha$, and (f) radius of curvature $r$ for the rounded corners.
The geometrical parameters of the nanobowtie are $w=80$ nm, $h=30$ nm, $d=4$ nm, $\alpha=60^{\circ}$, and $r=2$ nm, unless otherwise stated.
}
\label{figs2}
\end{figure}

\begin{figure}[!ht]
\centering
\includegraphics[scale=0.7]{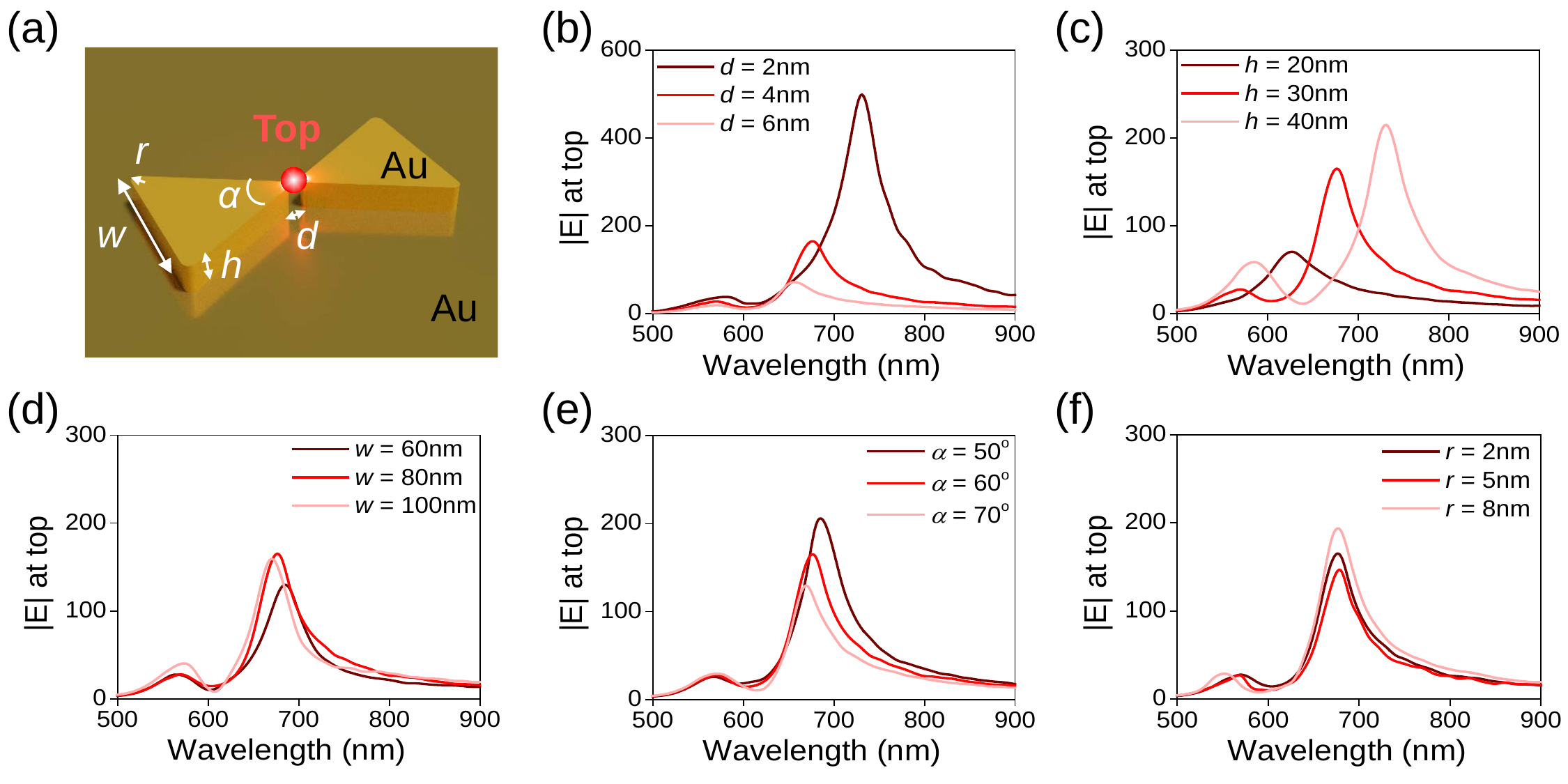}\\
\caption{
\textbf{Dependence of the field enhancement on geometry.}
(a) Schematic of the bowtie-on-Au cavity, with the field enhancement $|E|$ at the gap center of the bowtie upper surface tracked. (b-f) Variation of the field enhancement as a function of (b) gap size $d$, (c) height $h$, (d) width $w$, (e) apex angle $\alpha$, and (f) radius of curvature $r$ for the rounded corners.
The geometrical parameters of the nanobowtie are $w=80$ nm, $h=30$ nm, $d=4$ nm, $\alpha=60^{\circ}$, and $r=2$ nm, unless otherwise stated.
}
\label{figs2_E}
\end{figure}

The optical response of the bowtie is dependent not only on its dielectric environment, but also its geometry. As an illustrative example, we present here this dependence for the bowtie-on-Au device. Figure \ref{figs2} displays the variation of the bowtie absorption spectrum as a function of several key geometrical parameters, including the gap size $d$, height $h$, width $w$, apex angle $\alpha$ and radius of curvature $r$ of the rounded corners. It can be seen that the spectral response is extremely sensitive to $d$ and $h$, moderately so to $w$ and $\alpha$, and only weakly dependent on $r$ for the selected range of values. Here, we focus particularly on the resonance at longer wavelength (designated by a red arrow in each case), which is the same mode shown in Fig. 1(a) of the main text. In addition to spectral shifts and changes in the peak absorption rate, the corresponding field enhancement $|E|$ at the hotspot (located in the center of the gap at the upper surface of the bowtie in the case of a Au substrate) also changes accordingly. As shown in Fig. \ref{figs2_E}, this field enhancement $|E|$ is sensitive to all of these parameters, and generally rises with decreasing $d$, increasing $h$, increasing $w$, decreasing $\alpha$, and decreasing $r$. 
It should be underlined that the field enhancement $|E|$ at the gap center is typically smaller than the global maximum near-field enhancement attainable through the antenna mode of the bowtie-on-Au system, which is realized in greater proximity to the Au surface itself. In Fig. \ref{figs2_E}(f) particularly, $|E|$ for $r=8$ nm exceeds that for $r=2$ nm, since the intense field for $r=2$ nm is tightly confined at the Au surface, rather than at the gap center.

\clearpage
\section*{\large S3. Higher-order Non-radiative Mode of the Bowtie-on-Au Cavity}

\begin{figure}[!ht]
\centering
\includegraphics[scale=0.6]{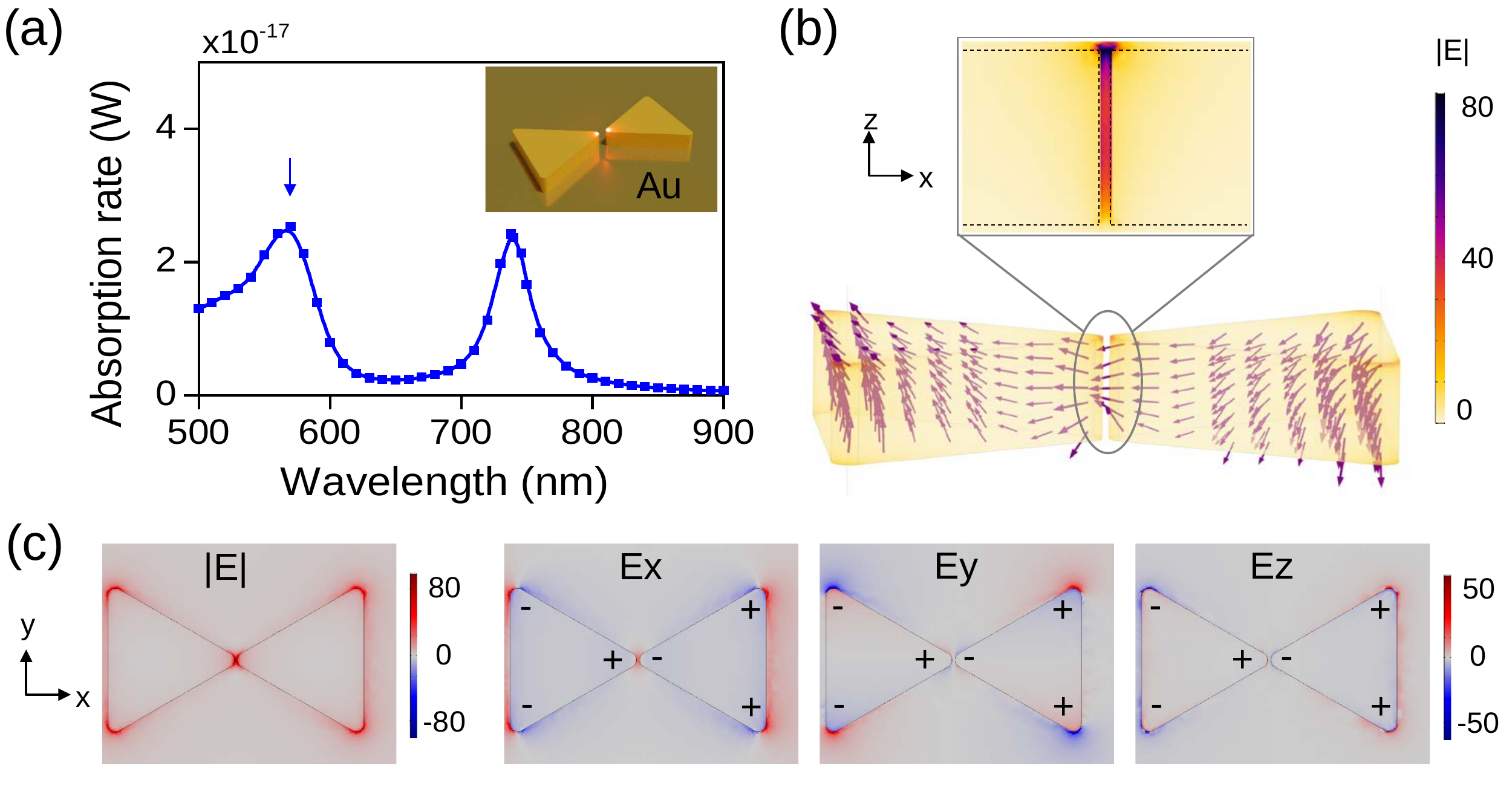}\\
\caption{
\textbf{Higher-order non-radiative mode of the bowtie-on-Au cavity.}
(a) Absorption spectrum of the nanobowtie. (b) Three-dimensional normalized electric near-field profile of the nanobowtie (color map) at 570 nm, with superimposed, vectorial, normalized electric field distribution (or charge flow) inside the device (purple arrows). The arrow direction indicates that of the electric field, while the arrow length is proportional to its magnitude in a logarithmic scale. Inset: the normalized electric near-field profile in the gap region (side view), showing the elevated hotspot. The dashed lines indicate the geometric boundaries of the nanobowtie. (c) Distributions of the normalized electric field magnitude and components at the upper surface of the nanobowtie (top view), where the superimposed $+$ and $-$ signs indicate the charge distribution polarities. The geometrical parameters of the nanobowtie are $w=100$ nm, $h=30$ nm, $d=2$ nm, $\alpha=60^{\circ}$, and $r=5$ nm.
}
\label{figs3}
\end{figure}

The bowtie-on-Au cavity has multiple resonances as suggested by the absorption spectrum and decay rate in Fig. 1(a) of the main text. Aside from the radiative mode around 738 nm discussed extensively in the main text, there is also a shorter-wavelength resonance around 570 nm, indicated by the arrow in Fig. \ref{figs3}(a). It corresponds to a higher-order mode that is non-radiative in character. In Fig. \ref{figs3}(b), the corresponding three-dimensional, normalized electric near-field profile is displayed, together with the charge flow inside the bowtie. Similar to the radiative mode, its associated hotspot is also located towards the top of the bowtie gap, but with a much weaker field enhancement ($|E|\sim$ 80 at 570 nm compared to $|E|\sim$ 480 at 738 nm). A second notable difference from the radiative mode studied in the main text is that this non-radiative mode shows a greater contribution to the electric near-field from the exterior corners of the bowtie, as suggested by the more prominent arrows towards them in Fig. \ref{figs3}(b) compared to those for the radiative mode [see Fig. 2(a) of the main text]. The same observation can also be made from the normalized near-field distributions pertaining to the electric field magnitude and components in Fig. \ref{figs3}(c), where the field enhancement $|E|$ at the exterior corners of the bowtie is found to be comparable to that in the gap. The distributions for $E_\mathrm{x}$, $E_\mathrm{y}$ and $E_\mathrm{z}$ around the exterior corners of the bowtie also correspond to the accumulation of positive/negative charges, as represented by the superimposed polarities in Fig. \ref{figs3}(c).

\clearpage
\section*{\large S4. Different Mechanisms for Hotspot Formation using Au and Si Substrates}

\begin{figure}[!ht]
\centering
\includegraphics[scale=0.6]{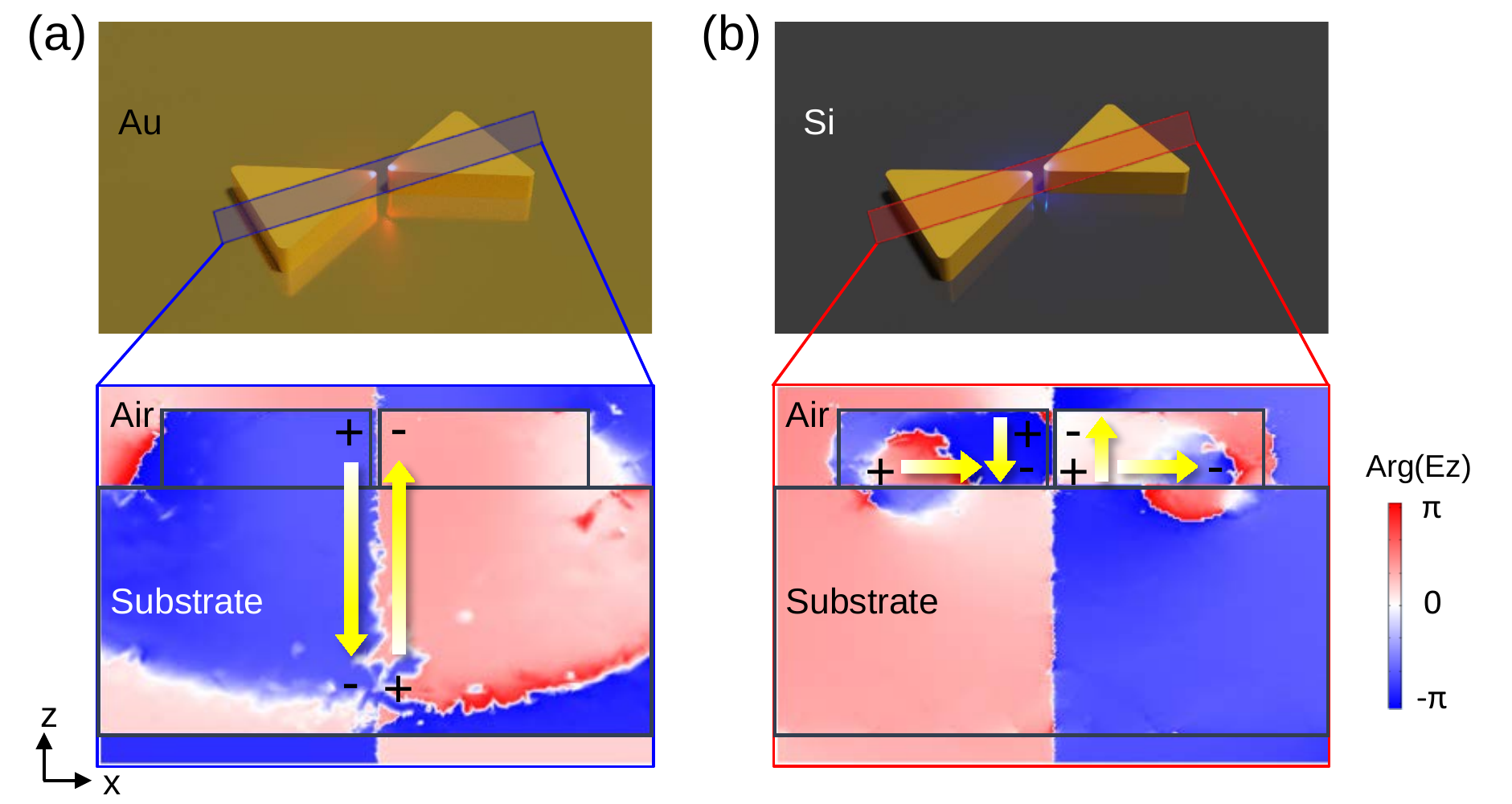}\\
\caption{
\textbf{Phase distributions for the antenna modes.}
(a-b) Schematics (top) and cross-sectional phase distributions of the $E_\mathrm{z}$ component (bottom) for (a) the bowtie-on-Au cavity at a wavelength of 738 nm and (b) the bowtie-on-Si cavity at a wavelength of 678 nm, respectively.
}
\label{figs4}
\end{figure}

Although the locations of the hotspots and corresponding field enhancements are similar in the cases of the bowtie-on-Au and bowtie-on-Si cavities (see Fig. 1 of the main text), the mechanisms behind their formation are different. This can be appreciated from the cross-sectional phase profiles of the $E_\mathrm{z}$ component shown in Fig. \ref{figs4}. Here, the red and blue regions indicate the $E_\mathrm{z}$ component with positive and negative phases, respectively. Therefore, the interface where the phase changes sign signifies an accumulation of surface charges with opposite polarities. By analyzing the sign of the $E_\mathrm{z}$ component in this way, we can denote the charge distribution polarities as shown in Fig. \ref{figs4}, and the corresponding charge flow (bright yellow arrows for illustration purposes only). For the bowtie-on-Au cavity in Fig. \ref{figs4}(a), the elevated hotspot originates from the coupling between the bowtie and its ``mirror image" (\textit{i.e.}, the surface charges towards the top of the bowtie gap and the corresponding image charges in the Au substrate). For the bowtie-on-Si cavity in Fig. \ref{figs4}(b), the charge is accumulated at the bowtie surfaces only (negligible in Si), which is reasonable. Furthermore, the latter mode is a higher-order one due to the large refractive index of Si, while in a similar wavelength range it is a bonding dimer plasmon mode for a bowtie-on-glass cavity (of the same physical dimensions).

\clearpage
\section*{\large S5. Two-level Maxwell-Bloch Simulations}

\subsection{\small Maxwell-Bloch Methodology}

To describe the dynamical interaction of an ensemble of quantum emitters with the plasmonic modes of each bowtie-on-substrate system considered in Sections IV and V of the main text, we employ a semiclassical Maxwell-Bloch methodology, which combines finite-difference time-domain (FDTD) modelling of the bowtie electromagnetic response with a quantum-mechanical, density-matrix treatment of the emitters.

The FDTD method \cite{TafloveHagnessbook} enables an accurate and computationally efficient numerical solution of Maxwell's equations in the form
\begin{equation}
    \mu_{0}\frac{\partial {\bf H}}{\partial t} = -\nabla\times{\bf E}, \label{ME1}
\end{equation}
\begin{equation}
    \epsilon_{0}\epsilon_{\textrm{b}}\frac{\partial {\bf E}}{\partial t} = \nabla\times{\bf H} - \frac{\partial {\bf P}}{\partial t}, \label{ME2}
\end{equation}
for the electric ${\bf E}({\bf r},t)$ and magnetic ${\bf H}({\bf r},t)$ fields. Here, $\epsilon_{0}$ and $\mu_{0}$ are the permittivity and permeability of vacuum respectively, $\epsilon_{\textrm{b}}({\bf r})$ is the dielectric function of any inactive media present in the system (such as metallic or dielectric substrates) and ${\bf P}({\bf r},t)$ is the macroscopic polarization of the active material.\ We assume non-magnetic media throughout.

To describe in a self-consistent fashion the interaction of such fields with an active layer of emitters, we couple Maxwell's equations with a simple optical Bloch model \cite{Boydbook,Carmichaelbook}. In this Maxwell-Bloch framework, each emitter is treated as an independent, quantum two-level system whereas the electromagnetic field is assumed classical. Denoting the ground and excited states of an individual emitter by $|1\rangle$ and $|2\rangle$ respectively, the Hamiltonian describing its interaction with an external field is given by
\begin{equation}
  \hat{H} = \hbar\omega_\mathrm{e} \hat{\sigma}^\dag\hat{\sigma}-\hat{\pmb\mu}\cdot {\bf E} = \hbar \omega_\mathrm{e} \hat{\sigma}^\dag\hat{\sigma}-{\pmb \mu} \cdot {\bf E} ({\hat{\sigma}}^\dag+\hat{\sigma}), \nonumber
\end{equation}
where $\omega_\mathrm{e}$ is the transition frequency, $\hat{\sigma}=|1\rangle\langle2|$ and $\hat{\sigma}^\dag=|2\rangle\langle1|$ are the lowering and raising operators of the system respectively, and ${\pmb \mu}=\langle 2|\hat{\pmb \mu}|1\rangle$ is a transition dipole matrix element.
A general state of the two-level system can be written as a linear combination of the ground and excited states,
\begin{equation}
  |\psi\rangle = c_1|1\rangle+c_2|2\rangle,
\end{equation}
and correspondingly, the density matrix for the two-level system is
\begin{equation}
\label{densitymatrix}
  \hat{\rho}
  =|\psi\rangle\langle \psi|. 
\end{equation}
    
In this work, we regard each two-level emitter as an open quantum system which, in addition to its dipole interaction with an external field, experiences relaxation and dephasing. The temporal evolution of the density matrix Eq. (\ref{densitymatrix}) under these conditions can be described by means of the following quantum master equation in Lindblad form \cite{Carmichaelbook},
\begin{align}\label{master1}
  \frac{\partial\hat{\rho}}{\partial t} = & - \frac{i}{\hbar}\lbrack \hat{H},\hat{\rho}\rbrack+\frac{\gamma_\mathrm{r}}{2}(2\hat{\sigma}\hat{\rho}\hat{\sigma}^\dag-\hat{\sigma}^\dag\hat{\sigma}\hat{\rho}) +\frac{\gamma_\mathrm{p}}{2}(2\hat{\sigma}^\dag\hat{\rho}\hat{\sigma}-\hat{\sigma}\hat{\sigma}^\dag\hat{\rho}-\hat{\rho}\hat{\sigma}\hat{\sigma}^\dag) \nonumber \\
  & + \frac{\gamma_\mathrm{d}}{2}(\hat{\sigma}_z \hat{\rho} \hat{\sigma}_z-\hat{\rho}),
\end{align}
where $\hat{\sigma}_z=|2\rangle\langle2|-|1\rangle\langle1|$,  $\gamma_\mathrm{r}$ and $\gamma_\mathrm{p}$ are the incoherent relaxation and pumping rates respectively, and $\gamma_\mathrm{d}$ is the pure dephasing rate. Forming matrix elements of the left- and right-hand sides of Eq. (\ref{master1}) in a basis comprising the emitter ground and excited states, we find a system of coupled, nonlinear, partial differential equations for the density matrix elements $\rho_{ij}$,
\begin{equation}
 \frac{\partial\rho_{22}}{\partial t}= -\frac{\partial \rho_{11}}{\partial t}  
 =-\gamma\left(\rho_{22}-\rho^{SS}_{22}\right)-\frac{2}{\hbar}{\pmb \mu} \cdot {\bf E} \cdot \mathrm{Im}[\rho_{12}],
 \label{MBint1}
\end{equation}
\begin{equation}
 \frac{\partial\rho_{12}}{\partial t}= -\frac{\partial \rho^\ast_{21}}{\partial t} 
 =-(\Gamma_\mathrm{e}-i\omega_\mathrm{e})\rho_{12}+\frac{i}{\hbar}{\pmb \mu} \cdot {\bf E}(\rho_{22}- \rho_{11}),
  \label{MBint2}
\end{equation}
where $\gamma= \gamma_\mathrm{r}+\gamma_\mathrm{p}$,
$\Gamma_\mathrm{e}=\gamma_\mathrm{d}+\gamma/2$ is the total dephasing rate
and $\rho^{SS}_{22}= \gamma_\mathrm{p} / (\gamma_\mathrm{r}+\gamma_\mathrm{p})$ is the steady-state value of the excited-state population. We note that at room temperature and optical frequencies, $\gamma_\mathrm{p} \ll \gamma_\mathrm{r}$ and so $\rho^{SS}_{22} \ll 1$. As such, we neglect $\rho^{SS}_{22}$ in our numerical simulations.

For an ensemble of emitters distributed with density $N_\mathrm{d}$,
we define the ground- and excited-state population densities as $N_{1}=N_\mathrm{d}\rho_{11}$ and $N_{2}=N_\mathrm{d}\rho_{22}$ respectively, and calculate the macroscopic polarization via

\begin{equation}\label{PolNd}
 {\bf P} = N_\mathrm{d}\mathrm{Tr}(\hat{\rho}\hat{\pmb \mu})
 = N_\mathrm{d}{\pmb\mu} (\rho_{12}+\rho_{21})=2N_\mathrm{d} {\pmb\mu} \mathrm{Re}[\rho_{12}].
\end{equation}

By separating Eq. (\ref{MBint2}) according to the real and imaginary parts of the off-diagonal density matrix elements, and using Eq. (\ref{PolNd}), we can rewrite the coupled system in terms of the macroscopic polarization ${\bf P}$,

\begin{equation}
 \frac{\partial^2 \bf P}{\partial t^2}+2\Gamma_\mathrm{e} \frac{\partial {\bf P}}{\partial t}+\left(\Gamma_\mathrm{e}^2+\omega^2_\mathrm{e}\right) {\bf P} = -\frac{2\omega_\mathrm{e} }{\hbar} {\pmb \mu} ({\pmb \mu} \cdot {\bf E})(N_{2}-N_{1}), 
 \label{MBpol}
\end{equation}

\begin{equation}
\frac{\partial N_{2}}{\partial t} = -\frac{\partial N_{1}}{\partial t}=-\gamma N_{2} +\frac{1}{\hbar \omega_\mathrm{e}}\left(\frac{\partial \bf P}{\partial t}+\Gamma_\mathrm{e} {\bf P}\right)\cdot {\bf E}.
\label{MBdiag}
\end{equation}

In this work, we solve numerically the system comprising Maxwell's equations Eqs. (\ref{ME1}-\ref{ME2}) and the optical Bloch equations Eqs. (\ref{MBpol}-\ref{MBdiag}) in a completely self-consistent fashion, where the local electric field (computed using the FDTD method) drives a polarization response from the quantum emitters (determined by solving the optical Bloch equations), which in turn couples back to the field.

\subsection{\small Single-emitter Dynamics}

In addition to a uniform sheet of two-level quantum emitters, we have also reported Maxwell-Bloch simulations for a single-cell emitter coupled to the antenna modes of bowtie-on-Au and bowtie-on-Si cavities in Sections IV and V. A well-known complication of such self-consistent numerical simulations is the need to carefully treat the divergent nature of the in-phase self-field of the emitter, which can give rise to nonphysical frequency shifts and becomes particularly important for the larger dipole moments (up to 20 D) treated in this work \cite{DeinegaSeideman2014,Schelew2017}. To that end, we follow the self-interaction correction scheme proposed by Schelew {\it et al.} \cite{Schelew2017}, for which more detailed analysis and discussion can be found in their work. 

Assuming that the emitter occupies a single grid cell of size $\Delta x$, the total electric field entering Eqs. (\ref{MBpol}-\ref{MBdiag}) is corrected by subtracting a numerically divergent contribution ${\bf E}^{\textrm{div}}$, given by
\begin{equation}
   {\bf E}^{\textrm{div}} = - \frac{\bf P}{3 \epsilon_0 \epsilon_B}\left[ 1 + f(\Delta x)\right], \nonumber
\end{equation}
where $\epsilon_{B}$ is the dielectric constant of the embedding medium (here $\epsilon_{B} = 1$) and
\begin{equation}
   f(\Delta x) = -\left(\frac{3}{4\pi}\right)^{2/3}\left(\frac{1.15\omega_\mathrm{e} {\Delta x} \sqrt{\epsilon_B}}{c}\right)^{2}.
\end{equation}

\subsection{\small Numerical Simulations}

Our spatiotemporal simulations of the coupled electromagnetic response and polarization dynamics of the plexcitonic systems of interest are performed using the photonic simulation software package Lumerical FDTD \cite{Lumericalwebsite}.
The bowtie structure is placed on an infinite Au or Si substrate, and open boundaries are simulated by means of perfectly matched layers. Our FDTD calculations (which follow a standard Yee algorithm \cite{Yee1966}) employ a 2 nm mesh size in a region of minimum dimensions 250 nm $\times$ 150 nm $\times$ $(h+16)$ nm containing the bowtie structure, which is progressively refined to a mesh size of 0.5 nm in a 12 nm $\times$ 12 nm $\times$ $(h+8)$ nm region around the center of the bowtie gap. The use of such a small mesh size ensures an accurate description of the spatial variations in the plasmonic near-field within the bowtie nanocavity. Regarding the active quantum matter, we treat both a layer of two-level emitters as well as a single such emitter interacting with the antenna modes of the bowtie-on-Au and bowtie-on-Si devices. In the former case, the layer comprises an ensemble of emitters distributed with uniform density $N_\mathrm{d} = 1$ nm$^{-3}$ across the upper surface of the bowtie structure (i.e., spanning a number of grid cells), while in the latter, a single-cell emitter is positioned at the top of the gap, 0.5 nm away from the apex of one of the nanoprisms.
For the calculation of the quantities $\gamma$ and $\Gamma_\mathrm{e}$ in Eqs. (\ref{MBpol}-\ref{MBdiag}), we disregard thermal pumping (which is negligible at room temperature, $\gamma_\mathrm{p} \approx 0$) and set $\gamma_\mathrm{r} = 10^{9}$ rad/s (corresponding to a nanosecond excited-state lifetime) and $\gamma_\mathrm{d} = 10^{13}$ rad/s. Note that as a result, $\Gamma_\mathrm{e}\approx\gamma_\mathrm{d}$. The injection of an $x$-polarized, exciting plane wave into the simulation domain and the calculation of the scattering cross-section for each coupled system are performed using the well-known total-field scattered-field technique \cite{TafloveHagnessbook}.

\clearpage
\section*{\large S6. Scattering Spectrum in the Single-emitter Strong Coupling Regime}

\begin{figure}[!ht]
\centering
\includegraphics[scale=0.7]{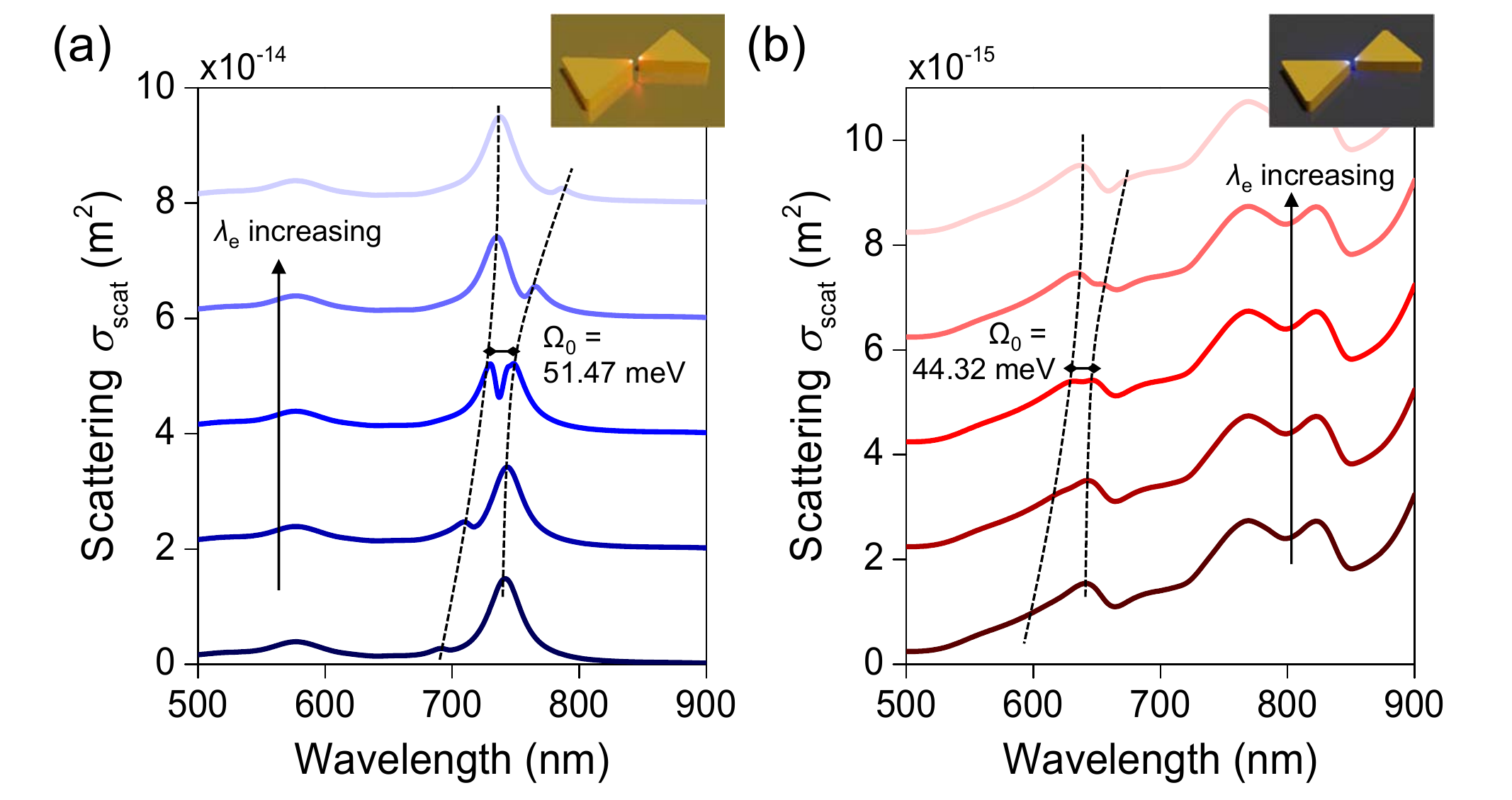}\\
\caption{
\textbf{Strong coupling of the bowtie-on-Au and bowtie-on-Si cavity antenna modes with a single quantum emitter.}
The scattering spectra of the plexcitonic system with the emitter transition wavelength swept for (a) the bowtie-on-Au and (b) the bowtie-on-Si cavities. Dashed black lines trace the anticrossing behaviour and Rabi splitting $\Omega_0$ in each case. 
The geometry of the nanobowtie is identical throughout with $w=100$ nm, $h=30$ nm, $d=2$ nm, $\alpha=60^{\circ}$, and $r=5$ nm.
}
\label{figs5}
\end{figure}

Complementary to our discussion on the strong coupling behaviour of emitter ensembles using the bowtie-on-Au and bowtie-on-Si cavities in Section IV of the main text, we consider here the scattering spectrum of each plexcitonic system with a single-cell emitter positioned at the nanogap hotspot. Figs. \ref{figs5}(a) and \ref{figs5}(b) present the scattering spectra for the bowtie-on-Au and bowtie-on-Si cavities each interacting with a single emitter. For both cavities, a dual-peak structure is observed with a clear anticrossing behaviour between the peaks (traced by the dashed black lines) as the emitter transition wavelength $\lambda_\mathrm{e}$ is swept through the cavity antenna-mode resonance. The observation of Rabi oscillations in the emitter polarization as a function of time (see Section IV of the main text) confirms that this spectral splitting manifests plexciton formation (\textit{i.e.}, the emitter and cavity are in the strong coupling regime). 

The corresponding cavity-emitter coupling strengths can be calculated according to a phenomenological, dissipative Jaynes-Cummings model \cite{xu2018quantum,Baranovetal2018review}. Here, the polaritonic peaks appear at frequencies of $\omega_{\pm}=\omega_\mathrm{c}\pm\sqrt{\Delta^2/4+g_0^2-(\Gamma_\mathrm{e}-\kappa_\mathrm{c})^2/4}$, where $\omega_{c}$ is the cavity mode frequency, $\Delta$ is the detuning between the emitter transition and plasmonic cavity mode, $g_0$ is the single-emitter coupling strength, $\Gamma_{\mathrm{e}}$ is the emitter decay rate and $\kappa_{\mathrm{c}}$ is the decay rate of the cavity mode. The minimal splitting, realized for zero detuning ($\Delta=0$), is given by $\Omega_0=\sqrt{4g_0^2-(\Gamma_\mathrm{e}-\kappa_\mathrm{c})^2}$ and indicated in Fig. {\ref{figs5}}. The antenna-mode decay rates $\kappa_\mathrm{c}^\mathrm{Au,Si}$ for the bowtie-on-Au and bowtie-on-Si cavities are determined as the half-widths at half-maximum of the corresponding peak features in the spectra of Figs. 1(a)iv and 1(b)iv in the main text, respectively. For the bowtie-on-Au cavity, using values of $\Gamma_\mathrm{e}=6.5$ meV and $\kappa_\mathrm{c}^\mathrm{Au}=42.08$ meV, the single-emitter coupling strength is found to be $g_0^\mathrm{Au}=31.28$ meV. Similarly, for the bowtie-on-Si cavity with $\kappa_\mathrm{c}^\mathrm{Si}=57.58$ meV, we find $g_0^\mathrm{Si}=33.81$ meV.


\clearpage
\section*{\large S7. Dependence of the Plexcitonic Rabi Splitting on the Bowtie Apex Angle}

\begin{figure}[!ht]
\centering
\includegraphics[scale=0.7]{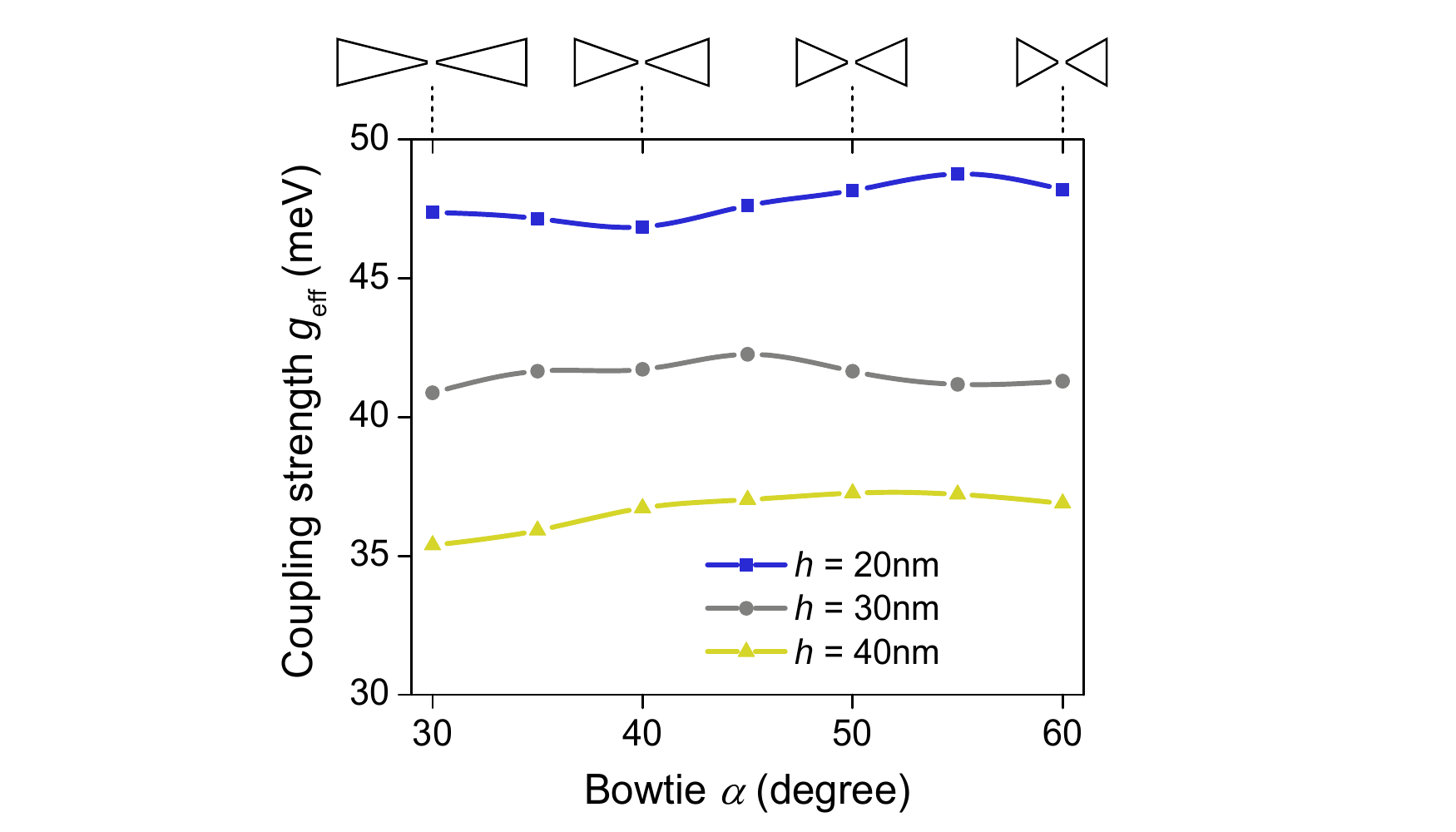}\\
\caption{
\textbf{Dependence of the effective coupling strength $g_\mathrm{eff}$ on the bowtie apex angle $\alpha$ for a sheet of quantum emitters placed on the bowtie-on-Au cavity.}
The data are extracted from the scattering spectra of the coupled system as the emitter transition is swept across the antenna-mode resonance, with different bowtie heights $h$. Insets: Schematics of the bowtie geometry as $\alpha$ is varied but $w$ is fixed. The geometrical parameters of the nanobowtie are $w=100$ nm, $h=30$ nm, $d=2$ nm, $\alpha=60^{\circ}$, and $r=5$ nm, unless otherwise stated.
}
\label{figs6}
\end{figure}

Besides the bowtie height, the apex angle $\alpha$ can also be used to tune the characteristics of many-emitter strong coupling. Figure \ref{figs6} shows the effective coupling strength $g_\mathrm{eff}$ as a function of $\alpha$ for the bowtie-on-Au cavity with a range of bowtie heights $h$, evidencing a weak dependence on the former for the investigated angular range, and a qualitatively similar dependence on the latter as in Fig. 4(b) of the main text. Although the coupling strength cannot be significantly improved by varying the bowtie $\alpha$, an optimal apex angle yielding the largest Rabi splitting can nevertheless be identified for each $h$ value. Furthermore, the apex angle offers additional freedom to tune the plasmonic resonance to spectrally match with different types of quantum emitter. Note that in these calculations, where the bowtie $\alpha$ is varied but $w$ is fixed, the size of the emitter layer is maximized only to cover the area of the bowtie with $\alpha=30^\circ$, and remains fixed otherwise.

\clearpage
\bibliography{reference}

